\documentclass[aps,pra,twocolumn,superscriptaddress,reprint,longbibliography]{revtex4-2}

\usepackage{graphicx}
\usepackage{dcolumn}
\usepackage{bm}
\usepackage[utf8]{inputenc}
\usepackage[T1]{fontenc}
\usepackage{mathptmx}
\usepackage{amssymb}
\usepackage[euler]{textgreek}
\usepackage{upgreek}
\usepackage{hyperref}
\usepackage{amsmath}
\usepackage[version-1-compatibility,per-mode=symbol]{siunitx}
\usepackage{color}
\usepackage{bm}
\usepackage{textcmds}
\usepackage{xspace}

\DeclareSIUnit[per-mode=symbol]{\intensity}{\watt\per\centi\meter\squared}
\DeclareSIUnit{\electrondensity}{e\ensuremath{^-}\per\centi\meter\cubed}
\newcommand{\micron}{\ensuremath{\upmu}\text{m}\xspace}

\begin{document}

	\title{Characterization of angularly resolved EUV emission from 2-\textmu m-wavelength laser-driven Sn plasmas using preformed liquid disk targets}
	\author{R Schupp}
	\affiliation{Advanced Research Center for Nanolithography, Science Park~106, 1098~XG Amsterdam, The Netherlands}
	\author{L Behnke}
	\affiliation{Advanced Research Center for Nanolithography, Science Park~106, 1098~XG Amsterdam, The Netherlands}
	\affiliation{Department of Physics and Astronomy, and LaserLaB, Vrije Universiteit, De Boelelaan 1081, 1081 HV Amsterdam, The Netherlands}
	\author{Z Bouza}
	\affiliation{Advanced Research Center for Nanolithography, Science Park~106, 1098~XG Amsterdam, The Netherlands}
	\affiliation{Department of Physics and Astronomy, and LaserLaB, Vrije Universiteit, De Boelelaan 1081, 1081 HV Amsterdam, The Netherlands}
	\author{Z Mazzotta}
	\affiliation{Advanced Research Center for Nanolithography, Science Park~106, 1098~XG Amsterdam, The Netherlands}
	\author{Y Mostafa}
	\affiliation{Advanced Research Center for Nanolithography, Science Park~106, 1098~XG Amsterdam, The Netherlands}
	\affiliation{Department of Physics and Astronomy, and LaserLaB, Vrije Universiteit, De Boelelaan 1081, 1081 HV Amsterdam, The Netherlands}
	\author{A Lassise}
	\affiliation{Advanced Research Center for Nanolithography, Science Park~106, 1098~XG Amsterdam, The Netherlands}
	\author{L Poirier}
	\affiliation{Advanced Research Center for Nanolithography, Science Park~106, 1098~XG Amsterdam, The Netherlands}
	\affiliation{Department of Physics and Astronomy, and LaserLaB, Vrije Universiteit, De Boelelaan 1081, 1081 HV Amsterdam, The Netherlands}
	\author{J Sheil}
	\affiliation{Advanced Research Center for Nanolithography, Science Park~106, 1098~XG Amsterdam, The Netherlands}
	\author{M Bayraktar}
	\affiliation{Industrial Focus Group XUV Optics, MESA+ Institute for Nanotechnology, University of Twente, Drienerlolaan 5, 7522 NB Enschede, The Netherlands}
	\author{W Ubachs}
	\affiliation{Advanced Research Center for Nanolithography, Science Park~106, 1098~XG Amsterdam, The Netherlands}
	\affiliation{Department of Physics and Astronomy, and LaserLaB, Vrije Universiteit, De Boelelaan 1081, 1081 HV Amsterdam, The Netherlands}
	\author{R Hoekstra}
	\affiliation{Advanced Research Center for Nanolithography, Science Park~106, 1098~XG Amsterdam, The Netherlands}
	\affiliation{Zernike Institute for Advanced Materials, University of Groningen, Nijenborgh 4, 9747 AG Groningen, The Netherlands}
	\author{O O Versolato}\email{o.versolato@arcnl.nl}
	\affiliation{Advanced Research Center for Nanolithography, Science Park~106, 1098~XG Amsterdam, The Netherlands}
	\affiliation{Department of Physics and Astronomy, and LaserLaB, Vrije Universiteit, De Boelelaan 1081, 1081 HV Amsterdam, The Netherlands}
	
	\date{\today}
	
\begin{abstract}
	The emission properties of tin plasmas, produced by the irradiation of preformed liquid tin targets by several-ns-long 2-\micron-wavelength laser pulses, are studied in the extreme ultraviolet (EUV) regime. In a two-pulse scheme, a pre-pulse laser is first used to deform tin microdroplets into thin, extended disks before the main (2\,\micron) pulse creates the EUV-emitting plasma. 
    Irradiating 30- to 300-\micron-diameter targets with 2-\micron laser pulses, we find that the efficiency in creating EUV light around 13.5\,nm follows the fraction of laser light that overlaps with the target. Next, the effects of a change in 2-\micron drive laser intensity (0.6--\SI{1.8E11}{\intensity}) and pulse duration (3.7--7.4\,ns) are studied.
    It is found that the angular dependence of the emission of light within a 2\% bandwidth around 13.5\,nm and within the backward 2$\pi$ hemisphere around the incoming laser beam is almost independent of intensity and duration of the 2-\micron drive laser. With increasing target diameter, the emission in this 2\% bandwidth becomes increasingly anisotropic, with a greater fraction of light being emitted into the hemisphere of the incoming laser beam. For direct comparison, a similar set of experiments is performed with a 1-\micron-wavelength drive laser. Emission spectra, recorded in a 5.5--25.5\,nm wavelength range, show significant self-absorption of light around 13.5\,nm in the 1-\micron case, while in the 2-\micron case only an opacity-related broadening of the spectral feature at 13.5\,nm is observed. 
    This work demonstrates the enhanced capabilities and performance of 2-\micron-driven plasmas produced from disk targets when compared to 1-\micron-driven plasmas, providing strong motivation for the use of 2-\micron lasers as drive lasers in future high-power sources of EUV light.
\end{abstract}

\maketitle
	
\section{Introduction}
	Laser-produced plasmas containing highly charged tin ions are the light source of choice for state-of-the-art extreme ultraviolet (EUV) lithography\,\cite{versolato2019physics,Purvis2018industrialization,Moore2018euv,Schafgans2015performance,mizoguchi2018high,OSullivan2015,Vinokhodov2016droplet,Harilal2011effect, Giovannini2014,Banine2011,Benschop2008}. Tin is used because no less than five of its charge states (Sn$^{10+}$--Sn$^{14+}$) strongly emit in a narrow band around 13.5\,nm \cite{Azarov1993,Churilov2006SnIX--SnXII,Churilov2006SnVIII,Churilov2006SnXIII--XV,Ryabtsev2008SnXIV,Tolstikhina2006ATOMICDATA,DArcy2009a,Ohashi2010,Colgan2017,Torretti2017,Scheers2020} that matches the peak reflectivity of available multilayer optics\,\cite{Bajt2002,Huang2017}. The tin ions are bred in a hot ($\sim$30--60\,eV) and dense ($10^{19}$--$10^{21}\,\electrondensity$) plasma. Starting from mass-limited tin-microdroplets, a low intensity \textit{pre-pulse} (PP) deforms the droplets into a shape better suited for interaction with a second high-intensity \textit{main pulse} (MP), used to create the EUV-emitting plasma. This two-step process is crucial for reaching source efficiencies and power levels that allow for the industrial utilization of EUV lithography \cite{nishihara2008plasma,Basko2016,Fomenkov2017, Fujioka2008}.
	Currently, 10.6-\micron-wavelength CO$_2$ gas lasers are used to drive the plasma. Recent simulation studies however have drawn significant attention to the use of a 2-\micron main pulse, a wavelength at which high-power solid-state lasers may soon become available \cite{Langer2020litho}. Typical advantages of near- to mid-infrared solid state over gas lasers are their more compact build and higher wall-plug efficiency. Shorter drive laser wavelengths have the additional advantage of a higher coupling efficiency of laser light with the tin plasma. Given a higher critical electron density ($n_\mathrm{c}\propto\lambda^{-2}$), the shorter wavelength laser light is absorbed in regions with higher emitter and absorber density. This may benefit the obtainable source brightness but an associated increase in optical depth \cite{Schupp2019b,Schupp2020} leads to increased broadening of spectral features outside of the for EUV lithography relevant \emph{in-band} region, defined as a 2\% bandwidth centered at 13.5\,nm. This broadening may limit the obtainable conversion efficiency (CE) of laser energy into in-band radiation into the backward hemisphere towards the laser origin. 
    Limitations on the spectral performance of EUV sources imposed by optical depth have been studied in detail for plasmas driven by 1-\micron Nd:YAG lasers \cite{Schupp2019,Schupp2019b,Fujioka2005opacity}. A 2-\micron wavelength, between the widely investigated 1 and 10\,\micron, is an interesting candidate providing intermediate plasma densities and hence optical depth. In addition to 2-\micron systems based on difference frequency generation\,\cite{Arisholm2004,Behnke2021,Schupp2020}, high-power Big Aperture Thulium (BAT) laser systems, operating at 1.9-\micron wavelength, are currently under development \cite{Danson2019petawatt,Sistrunk2019}.
    
    Despite its importance for EUV lithography, literature on the production of in-band radiation from mass-limited, pre-pulse deformed targets remains rather scarce, with the majority of studies to-date focusing on $\lambda$\,=\,10\,\micron main pulses (see, e.g., \cite{Fujioka2008,Matsukuma2015,Higashiguchi2006,Nakamura2008}). A broad body of literature has investigated experimentally the emission from undeformed liquid tin droplet targets \cite{Giovannini2013,Giovannini2014,Schupp2019,Sequoia2008}, coated spheres \cite{Shimada2005} or solid-planar targets \cite{Ando2006optimum,Choi2000,Freeman2012laser,Morris2008Angular} using 1-\micron solid-state lasers. First results from experiments using 2-\micron laser light driving plasma from solid-planar tin targets \cite{Behnke2021} as well as from undeformed liquid tin droplet targets \cite{Schupp2020} have recently been presented. Both works demonstrate the potential of the 2-\micron drive wavelength, showing a doubling of the obtainable CE over the 1-\micron driver, to a 3\% level for the solid-planar tin target. At this 3\% level the overall conversion efficiency of wall plug power to in-band EUV may be at par with the current CO$_2$-driven industrial solutions. Literature on industrially-relevant mass-limited deformed targets interacting with either 1- or 2-\micron main pulse beams at high CE is however not yet available.
    
    In this publication, EUV light production from mass-limited tin targets, suitably shaped by a 1-\micron laser pre-pulse, is investigated using 1- and 2-\micron main pulse laser systems to drive the plasma. First, plasma produced with a 2-\micron laser beam is investigated for three different laser intensities and three laser pulse durations. For each laser parameter set a wide range of target diameters (30--300\,\micron) is investigated and the resulting angular distribution of in-band emission, the efficiency of converting laser light into in-band EUV radiation and overall EUV radiation, as well as the spectral performance of the plasma, are characterized. Secondly, plasmas generated by 1- and 2-\micron drive laser light are characterized and compared.

\section{Experiment}

\begin{figure}[tb]
	\centering
	\includegraphics[scale=0.5]{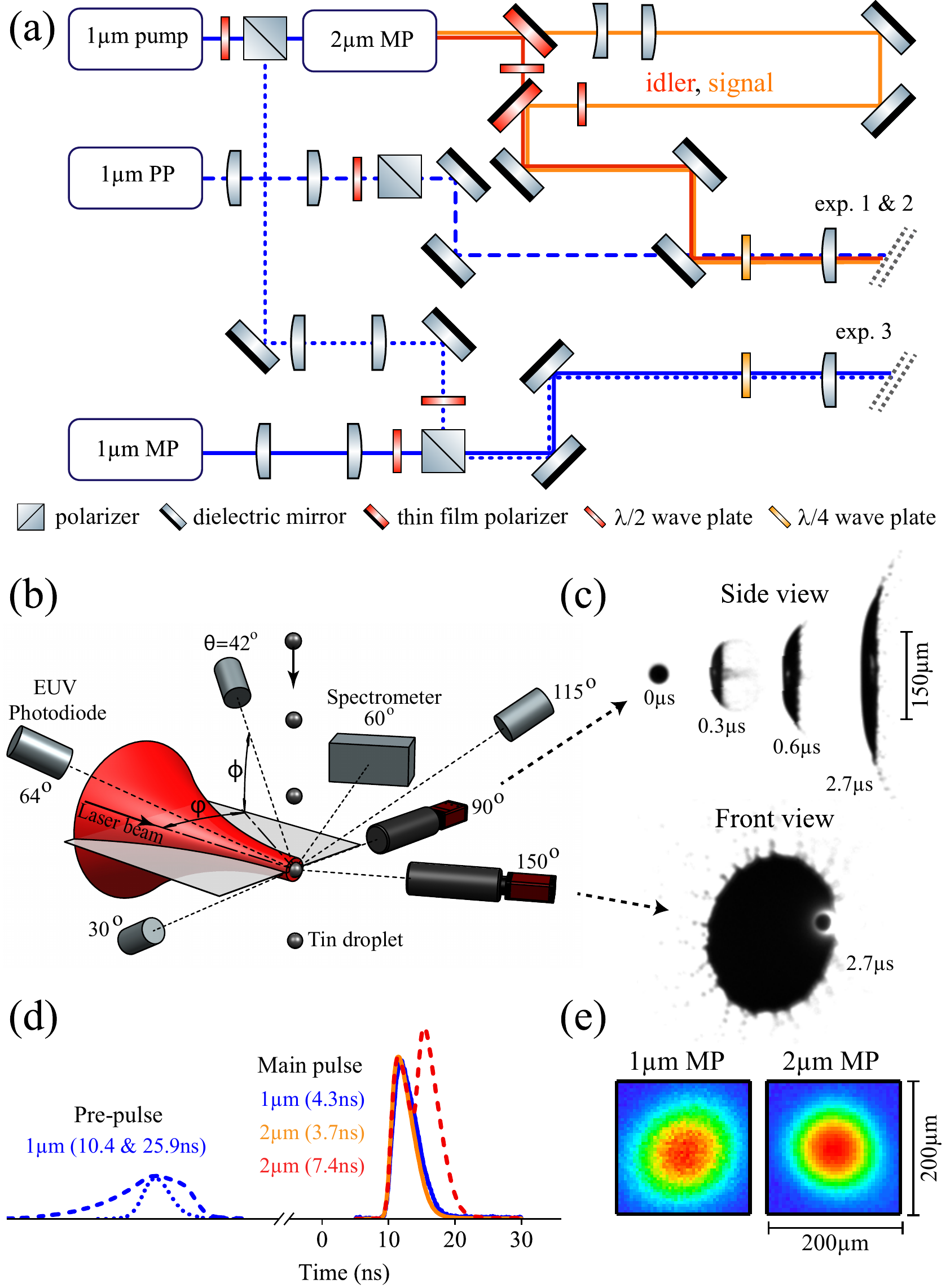}
	\caption{(a) Schematic representation of the laser beam setups for experiments 1 \& 2 and experiment 3 with pre-pulse (PP) and main pulse (MP) lasers shown. (b) Continuation from (a) showing the locations and angles of the detectors. All angles $\theta$ are calculated with respect to the laser-beam propagation axis, with $\cos \theta= \cos \phi \cos \varphi$. (d) Selection of front and side view shadowgraphs of the tin targets used for plasma generation, recorded by the two cameras indicated in (b). (d) Temporal and (e) spatial profiles of the laser beams. For more detail see text.}
	\label{ch5-fig:1}
\end{figure}

In the experiments, micrometer-sized liquid tin droplets are first irradiated with a relatively low intensity ($\sim$\SI{E9}{\intensity}), 1-\micron wavelength PP (see Fig.\,\allowbreak\ref{ch5-fig:1}(a)) from an Nd:YAG laser. The PP propels the droplets and deforms them into extended, disk-like targets of diameter $d_\mathrm{T}$\,\cite{Kurilovich2016,Kurilovich2018,Liu2020a,Liu2021a} with typical radial expansion speeds of $\sim$\SI{90}{\meter\per\second}. The target diameter is precisely controlled via the time delay between pre- and main pulse, which ranges 0--3\,\textmu s. 
In all measurements concerning 2-\micron main pulses, the spatial beam profile of this PP laser is Gaussian with a size of 120\,\micron at full width at half maximum (FWHM) and  a FWHM duration of 25.9\,ns (see Fig.\,\ref{ch5-fig:1}(d)). A constant PP energy of 8.4\,mJ is used throughout the experiments. We use circular polarization, to ensure that the produced targets are radially symmetric \cite{Liu2020a,Pinto2021}. The produced tin targets are observed under angles of \SI{90}{\degree} (side view) and \SI{150}{\degree} (front view) with respect to the laser beam using combinations of CCD cameras and long-distance microscopes (see Fig.\,\ref{ch5-fig:1}(b)). Temporal resolution is achieved by back-lighting the tin targets with spatially and temporally incoherent 560-nm-wavelength pulses of 6\,ns duration. Examples of typical targets are shown in Fig.\,\ref{ch5-fig:1}(c).

After a set time delay the targets are irradiated with high-intensity 2-\micron-wave\-length laser pulses. The pulses are produced in a master oscillator power amplifier (MOPA) \cite{Schupp2020,Behnke2021} that was built following the work of Arisholm \textit{et al.}\,\cite{Arisholm2004}. Signal and idler pulses having energies up to 180\,mJ each are produced at wavelengths of 1.9 and 2.1\,\micron, respectively. Depending on the precise experimental conditions, the pulse duration of the MOPA system can vary between 3.7 to 4.6\,ns, and the exact, measured pulse durations are stated for each measurement in the following. 

For the first set of experiments, the signal beam is removed via polarization optics and only the idler beam of the MOPA is focused onto the tin targets while scanning laser intensity and target size. The focal spot of the idler beam is Gaussian and has a FWHM of 106\,\micron (see Fig.\,\ref{ch5-fig:1}(e)).

In a second set of experiments, the pulse duration of the 2-\micron beam is varied between 3.7\,ns (idler only) and 7.4\,ns (idler and signal) at equal laser intensity. To achieve the longer pulse duration, the signal beam, which is separated from the idler via a thin film polarizer (TFP), is sent into an optical delay line of 1.2\,m length. The focusing conditions of the delayed signal beam are matched to those of the idler by adjustment of beam size and collimation via a telescope within the delay line, resulting in a Gaussian focal spot size of 106\,\micron FWHM for both beams. After collimation adjustment the beams are combined again with a second TFP. The signal energy is adjusted via a $\lambda/2$-waveplate before this second TFP. 

To enable a direct comparison of plasmas driven by 1- and 2-\micron-wavelength laser pulses, a separate and final set of experiments is conducted where the 1-\micron beam of a third laser system is used as MP (lower part of Fig.\,\ref{ch5-fig:1}(a)). This 1-\micron MP laser is a seeded Nd:YAG laser with arbitrary temporal pulse-shaping capabilities\,\cite{Meijer2017} that are used to reproduce the temporal profile of the 4.3\,ns and the 7.4\,ns 2-\micron beam. The pump laser of the MOPA is used for pre-pulsing in this third set of experiments. The PP parameters are tuned to obtain similar radial expansion speeds as in experiments 1 and 2 and no significant change in EUV emission is observed when using either PP.
Using telescopes, 1-\micron PP and MP are adjusted in size and collimation. For the MP a symmetric, 90-\micron-sized focal spot is achieved close to the dimensions of the 2-\micron MP. The PP has a spot size of 100\,\micron and a pulse duration of 10.4\,ns.
PP and MP are combined via a polarizing beam cube at which their pulse energy is adjusted via a preceding $\lambda/2$-waveplate.  Consecutively, the combined beams are steered onto the droplet, passing a $\lambda/4$-waveplate just before the final focusing element. The latter changes the polarization of the beams from linear to circular, ensuring a symmetric deformation of the tin droplets by the PP laser as before.  
Laser intensity is calculated with respect to its peak in space and time according to $I_\mathrm{L}=(2\sqrt{2\ln{2}/2\pi})^3 E_\mathrm{L}/ab t_\mathrm{p}$, where $E_\mathrm{L}$ is the laser energy, $a$ and $b$ are the major and minor axis of the slightly elliptical beam given as FWHM, and $t_\mathrm{p}$ is the pulse duration.  

The spectral emission from the plasma is recorded using a transmission grating spectrometer \cite{Bayraktar2016broadband}. The spectrometer is mounted at an angle of \SI{60}{\degree} with respect to the laser axis (see Fig.\,\ref{ch5-fig:1}(b)) and is operated with a 25\,\micron slit and a 10\,000\,lines/mm grating. Subsequent to recording the spectra with a CCD camera, a dark exposure is subtracted from the images in order to account for thermal and readout counts on the CCD. Next, spectra are corrected for second order diffraction from the grating, the grating's first order diffraction efficiency and the quantum efficiency of the CCD. 

The absolute amount of in-band radiation is measured with four photodiode assemblies installed under angles of $\theta = 30$, 42, 64 and \SI{115}{\degree} with respect to the target's surface normal. 
The photodiode assemblies use a multilayer mirror near normal incidence to reflect the in-band light onto the photodiode detectors. Remaining optical radiation is filtered by an EUV-transmissive Si/Zr coating on the photodiodes. 
Next, following the approach of Ref.\,\cite{Schupp2019} (and analogously to \cite{Ando2006optimum} and \cite{Yamaura2005}) the measured in-band energy values are corrected for the respective solid angle and fitted with the monotonous smooth function
\begin{equation}
\label{eq:anisotropy-fit}
f(\theta) = (\alpha-\beta)\cos\left(\theta/2\right)^\gamma+\beta,
\end{equation}
with amplitude $\alpha$, offset $\beta$ and power $\gamma$. We note that this function differs from the fit function ${\sim} \cos\left(\theta\right)^{\gamma}$ used in Schupp \emph{et al.} \cite{Schupp2019}. The adaption is needed to capture the emission under angles $\theta > 90^{\circ}$ also in cases $\gamma < 1$. The resulting fitted curves, using either fit function, are found to be indistinguishable over the backward hemisphere where $\theta < 90^{\circ}$. The integral amount of in-band radiation over the hemisphere around the incoming laser beam is then calculated by integration of Eq.\,\eqref{eq:anisotropy-fit}
\begin{equation}
\label{eq:anisotric-ce}
E_{\mathrm{IB},2\pi} = 2\pi \int_{0}^{\pi/2} f(\theta) \sin{\theta} d\theta.
\end{equation}
Using this value, CE is defined as $E_{\mathrm{IB},2\pi}/E_\mathrm{L}$. We further define an anisotropy factor of the in-band emission to gauge which fraction of in-band light is emitted into the backward hemisphere via $E_{\mathrm{IB},2\pi}/E_{\mathrm{IB},4\pi}$, where $E_{\mathrm{IB},4\pi}$ is obtained by changing the upper integration boundary in Eq.\,\ref{eq:anisotric-ce} to $\pi$, hence integrating over all angles. A large anisotropy factor is favorable for industrial application as light-collecting optics typically cover the backward hemisphere.

A second measure for source performance is the spectral purity (SP), defined as the ratio of in-band energy to the total EUV energy. All SP values provided in the following are calculated with respect to the spectral range of 5.5--\SI{25.5}{nm} measured under a 60 degree angle. 
A third measure, the radiative efficiency $\eta_\mathrm{rad}$ of the plasma is defined as the ratio of CE over SP following Ref.\,\cite{Schupp2019} and yields the total amount of EUV light emitted in the 5.5--\SI{25.5}{nm} wavelength band per incoming laser energy. 
In the current work, and in contrast to Ref.\,\cite{Schupp2019}, $\eta_\mathrm{rad}$ is defined using the values measured at 60$^\circ$, i.e., using the in-band EUV energy emitted per steradian under an angle of 60$^\circ$ (this value is subsequently multiplied with 2$\pi$), meaning $(E_{\mathrm{IB},60^\circ}/E_\mathrm{L})/\textrm{SP}$, instead of using CE that is defined over the entire laser-facing hemisphere.
This is undertaken because the spectra and hence the SP values are expected to show strong angular dependence for expanded targets \cite{Hayden2006,Morris2008Angular}, a dependence that is further expected to be influenced by the target diameter.

\section{EUV generation using 2-\texorpdfstring{$\boldsymbol{\upmu}$\lowercase{m}}{um} light on preformed targets}
\label{sec:2um}

\begin{figure*}[tbp]
	\centering
	\includegraphics[scale=1]{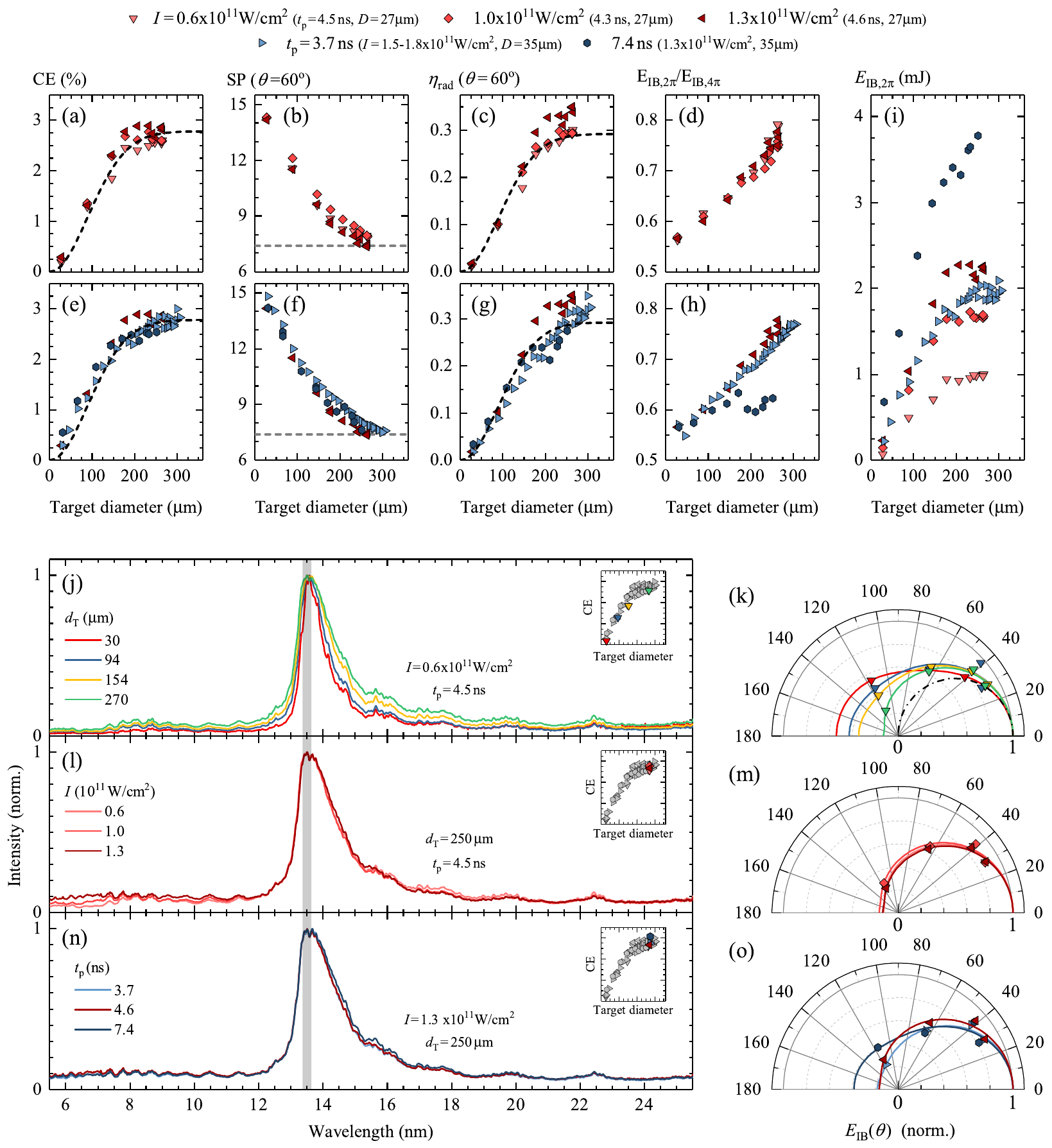}
	\caption{Results obtained with a 2-\micron drive laser wavelength. Panels (a),\,(e) depict CE; (b),\,(f) SP; (c),\,(g) $\eta_\mathrm{rad}$; (d),\,(h) anisotropy factor $E_{\mathrm{IB},2\pi}$/$E_{\mathrm{IB},4\pi}$; and (i) $E_{\mathrm{IB},2\pi}$, as function of target diameter, for three different intensities and laser pulse durations. SP and $\eta_\mathrm{rad}$ are obtained from measurements of the spectrum and in-band energy under a \SI{60}{\degree} angle. During the intensity scans the idler beam had a slightly longer pulse duration of 4.5\,ns (cf. 3.7\,ns during pulse duration scan) and the \SI{1.3e11}{\intensity} scan is shown in the pulse duration scan also. The dashed lines in panels (a), (c), (e), and (g) represent the results of a fit of the geometrical overlap of laser beam and target (EoT). The gray dashed lines in (b) and (f) indicate the SP values from measurements on planar-solid tin targets obtained from \cite{Behnke2021}. Also shown are normalized spectra versus (j) target size, (l) laser intensity and (n) pulse duration. The corresponding angular dependence of in-band emission, normalized at \SI{0}{\degree}, is shown in panels (k), (m) and (o). The solid lines depict fits of Eq.\,\eqref{eq:anisotropy-fit} to the data; the dash-dotted line in (k) indicates the $\cos(\theta)$ dependence of the Lambert cosine law.}
	\label{ch5-fig:2}
\end{figure*}

In the following, the influence of target diameter, laser intensity and laser pulse duration on CE, SP, $\eta_\mathrm{rad}$, $E_{\mathrm{IB},2\pi}$ and $E_{\mathrm{IB},2\pi}/E_{\mathrm{IB},4\pi}$ are investigated. Further, the spectral emission and angular distribution of the in-band emission are discussed. 

\subsection{Target diameter}
\label{subsec:2um-target-size}

To investigate the influence of target size, tin targets of various sizes as produced from pre-pulse-impacted 27-\micron-diameter droplets are irradiated with laser pulses having an intensity of \SI{0.6E11}{\intensity} and a pulse duration of 4.5\,ns (downward triangles in Fig.\,\ref{ch5-fig:2}). Data recorded using higher laser intensities and different pulse lengths are also shown but will be discussed in Sections\,\ref{subsec:laserintensity} and \ref{subsec:pulseduration}, respectively. Starting from an undeformed droplet target, CE has a low value of approximately $0.15\%$ that almost linearly increases with target diameter (see Fig.\,\ref{ch5-fig:2}(a)). At diameters of 170\,\micron and above, CE starts to plateau at a value of approximately 2.6\%. This plateau sets in where the target is roughly a factor of 1.5--2 larger than the FWHM size of the beam. A similar result was observed for 1-\micron-driven plasmas from tin coated glass spheres by Yuspeh \textit{et al.} \cite{Yuspeh2008optimization}. 

An increase in CE with target diameter is expected as an enlarged target diameter increases the geometrical overlap of laser beam and target. This overlap fraction is referred to as energy-on-target (EoT, following the definition in Refs.\,\cite{Kurilovich2016} and \cite{Kurilovich2018}). We note that there is a near complete absorption of laser light by the efficient inverse bremsstrahlung mechanism for this wavelength laser light at the here-relevant intensities \cite{Kurilovich2018}, where laser light geometrically overlaps with the target. For a circular spot, EoT is given by the function $A(1-2^{-d_\mathrm{T}^2/d_\mathrm{L}^2})$ which is an excellent approximation for the current focal spot shapes (cf. Fig.\,\ref{ch5-fig:1}(e)), where $d_\mathrm{L} = (a+b)/2$ and $a$ and $b$ are similar to within a few percent. The free fit factor $A$, the amplitude of the EoT curve, is obtained from a global fit to the asymptotic CE value of all data presented in Fig.\,\ref{ch5-fig:2}(a) and \ref{ch5-fig:2}(e). The observed dependence of CE on target size is seen to be well approximated by the geometric overlap function (dashed line in Fig.\,\ref{ch5-fig:2}(a)).

All CE values are calculated from angular-resolved measurements of the plas\-ma's in-band emission and all data points represent the mean over 300 individual laser shots. The angular dependence of the measured in-band energies (see Fig.\,\ref{ch5-fig:2}(k)) is first fitted with Eq.\,\eqref{eq:anisotropy-fit} and the CE value is subsequently calculated using the integration result from Eq.\,\eqref{eq:anisotric-ce}. For better visibility, all curves in Fig.\,\ref{ch5-fig:2}(k) are normalized at \SI{0}{\degree}. Eq.\,\eqref{eq:anisotropy-fit} is seen to accurately describe the angular dependence of the in-band EUV emission. The increase of in-band light observed at a target diameter of 94\,\micron and under an angle of \SI{42}{\degree} is however not fully captured by the fit function and this data point may in fact be an outlier. 

The reduced emission in the propagation direction of the laser at \SI{180}{\degree} with increasing target size may be caused by the one-sided heating of the expanded tin target by the laser beam. Plasma emission in this direction is shielded by the target (which is still thought to be present in liquid form during the laser pulse) and only plasma formed on the edge of the target presumably contributes to in-band emission under angles larger than \SI{90}{\degree}. 
The dashed line in Fig.\,\ref{ch5-fig:2}(k) shows Lambert's cosine law for the angular emission from a planar surface. With increasing target diameter the in-band emission starts to converge to this $\cos(\theta)$ dependence, however without fully reaching it. This might be expected because of the 3-dimensional extent of the plasma, causing departure of the emission characteristics from a Lambert-type distribution. Furthermore, emission will always exist at angles larger than \SI{90}{\degree} from the edge of the plasma unless $d_\mathrm{T}\gg d_\mathrm{L}$.

Next, the angular in-band distribution is quantified via an anisotropy factor, defined as the ratio of $E_{\mathrm{IB},2\pi}$/$E_{\mathrm{IB},4\pi}$, meaning the fraction of \textit{all} in-band energy that is emitted into the backward hemisphere of the incoming laser beam, relevant for EUV lithography applications. The anisotropy factor starts at a value of 0.57 for the droplet target, where 0.5 indicates an equal split between both hemispheres (see Fig.\,\ref{ch5-fig:2}(d)). The anisotropy factor then linearly increases with target diameter up to a value of 0.79, meaning that an ever-larger fraction of the in-band EUV is emitted into the backward hemisphere. 

The in-band energy emitted in the 2$\pi$ hemisphere towards the laser (obtained from multiplying the CE with the laser pulse energy) is shown in Fig.\,\ref{ch5-fig:2}(i). Naturally, the same trends are observed as in the case of CE. A maximum of $\sim$1\,mJ of in-band EUV energy is produced (per laser shot) for this lowest intensity case. 

Fig.\,\ref{ch5-fig:2}(j) depicts spectra from the lowest intensity case for a succession of four target diameters. For visibility and ease of comparison the spectra are normalized to their respective maximum values. Similar to previously published work, where a 2-\micron laser was used to produce plasma from spherical, undisturbed droplet targets \cite{Schupp2020}, all spectra show a strong emission feature at 13.5\,nm originating from transitions in Sn$^{8+}$--Sn$^{14+}$ ions\,\cite{OSullivan2015}. Transitions in these charge states further lead to the radiation observed in the 6--12\,nm wavelength region \cite{Torretti2018}.
With increasing target diameter, the feature at 13.5\,nm widens significantly from  0.8 to 1.5\,nm (FWHM). These are values between those of 1-\micron-driven low-density plasmas from planar SnO$_2$ targets of 0.5--1.5\,nm \cite{OSullivan1994,Choi2000} and plasmas from solid-planar tin targets of typically 2--3\,nm \cite{Hayden2006}. In Ref.\,\cite{Schupp2019} a FWHM of 0.9\,nm was reached in the case of a 1-\micron-driven plasma on a tin microdroplet target. 10-\micron laser-driven tin plasmas typically have a narrower 13.5-nm feature, reaching 0.6\,nm (FWHM)\,\cite{Kerkhof2020}. The 2-\micron-driven plasma is thus seen to produce spectra with spectral widths typically in between those produced with 1- and 10-\micron laser-driven pure tin plasmas. 

In contrast to the scaling of CE, SP has its maximum at the smallest target diameter. The highest value shown in Fig.\,\ref{ch5-fig:2}(b) is 11.5\% at a 90\,\micron diameter down from a maximum value of $\sim$14\% for the case of an undeformed droplet. SP for the undisturbed droplet target for this lowest intensity case is omitted from Fig.\,\ref{ch5-fig:2}(b) due to the low signal-to-noise ratio in the recorded spectrum. With increasing target diameter SP is observed to decrease monotonously, an effect previously observed for 1-\micron beams on droplet targets \cite{Schupp2019}. Above 200\,\micron diameter the decrease in SP plateaus towards a value of 7.5\% at a 260\,\micron diameter. These SP values are consistent with Ref.\,\cite{Behnke2021}, where the SP for planar-solid tin target plasmas was measured at 7.4\% (dashed line in panel (b)).

Using these SP values as input, a monotonic increase of $\eta_\mathrm{rad}$ with target diameter is observed up to a value of $\sim$0.3 at a 260\,\micron target diameter (cf. Fig.\,\ref{ch5-fig:2}(c)). Also indicated is the dependence of EoT on target size, now with an amplitude fit value of 0.29 (dashed curve). The here measured maximum radiative efficiency at \SI{60}{\degree} is slightly lower than a value of approximately 0.4 measured for an extended solid-planar target \cite{Behnke2021}. This difference originates from the slightly higher in-band emission at \SI{60}{\degree} measured on the planar target and may be explained by the smaller beam-spot size in the planar target case that allows for the in-band emission to escape more freely. Over the entire range of target diameters, the radiative efficiency qualitatively follows the EoT trend. 

At this point, it is worthwhile to consider the fact that both overall CE and $\eta_\mathrm{rad}$ qualitatively follow the EoT curve, describing the increasing geometrical overlap between an enlarging target and laser beam spot. However, we also note that the SP, which serves as input for calculating $\eta_\mathrm{rad}$, monotonically decreases with increasing target size. 
This apparent contradiction is resolved when considering the angular dependence of the plasma emission, with the increasing anisotropy factor serving to offset the decrease in SP with increasing target size (cf. Fig.\,\ref{ch5-fig:2}(b) and \ref{ch5-fig:2}(d)).

\subsection{Laser intensity}
\label{subsec:laserintensity}

Next, target diameter scans are performed for two higher laser intensities of 1.0 and \SI{1.3e11}{\intensity}. The CE curves for all three laser intensities show a trend very similar to the \SI{0.6e11}{\intensity} intensity case presented in Sec.\,\ref{subsec:2um-target-size} (see Fig.\,\ref{ch5-fig:2}(a)). For all target diameters, CE values are found to be slightly higher at higher laser intensities.
The maximum CE value at large target diameters increases modestly from 2.6 to 2.9\%. The angular distribution of in-band emission is observed to be independent of laser intensity within the scanned range (cf. Fig.\,\allowbreak\ref{ch5-fig:2}(m)). 
Given the constant maximum CE values, the amount of in-band radiation increases linearly with increasing laser intensity, with up to 2.3\,mJ of in-band light obtained for the highest intensity case (see Fig.\,\ref{ch5-fig:2}(i)).  

In Fig.\,\ref{ch5-fig:2}(l) normalized spectra for a target diameter of 250\,\micron{} are shown. For all spectra, the 8--25\,nm region looks remarkably alike. The most prominent difference between the spectra is seen in the 5--8\,nm region where the amount of radiation is observed to increase with increasing laser intensity. This emission could stem from charge states Sn$^{14+}$ (and above) as well as from an increased fraction of light being emitted from electronic states having higher excitation energies. 
The corresponding SP values are shown in Fig.\,\ref{ch5-fig:2}(b). All laser intensity cases follow the same trend. The maximum SP value for the two higher intensities is 14.5\% and is reached at the smallest target diameter, i.e., the undeformed droplet target. SP steadily decreases with increasing target size, and levels off at a value of up to 8\% depending on intensity. Even though the differences in SP are small, we do note that optimum SP values are observed for an intermediate intensity of \SI{1.0e11}{\intensity}.

Given the similar scaling of CE and SP with target diameter for the three laser intensities, the radiative efficiency in Fig.\,\ref{ch5-fig:2}(c) closely follows the \SI[allow-number-unit-breaks]{0.6e11}{\intensity} results, with a slightly higher value of 0.35 obtained for the highest intensity case. 

No significant changes are thus observed in the scaling of emission characteristics with target size when changing laser intensity and in-band EUV emission is seen to exhibit a linear dependence on laser energy.

\subsection{Pulse duration}
\label{subsec:pulseduration}

Lastly, the laser pulse duration is varied from 3.7 to 7.4\,ns (blue markers in Fig.\,\ref{ch5-fig:2}) by combining signal and idler beams while delaying the signal with respect to the idler pulse. To obtain identical focal spot conditions for both beams, the beam size of the signal beam is carefully matched to that of the idler beam by means of a telescope. The energy of both beams is individually controlled via combinations of $\lambda/2$-plates and TFPs. Signal and idler beams are set to have equal energies. The resulting temporal pulse profile is shown in Fig.\,\ref{ch5-fig:1}(d). Measurements are taken with 35-\micron-diameter droplets, slightly larger than those used during the laser intensity scan. We note that no dependence of CE on the initial droplet diameter is observed for target size scans performed for a range of droplet diameters from 19--45\,\micron diameter as long as the target thickness is sufficient to supply tin throughout the laser pulse. As an additional data set for the laser pulse duration we add the high-intensity (\SI{1.3E11}{\intensity}) scan from the previous Sec.\,\ref{subsec:laserintensity} that had slightly longer idler pulse duration of 4.6\,ns due to slightly different settings of the MOPA system. 

Again, CE is found to follow the same trend with increasing target diameter as described in Sec.\,\ref{subsec:2um-target-size}. A slightly higher maximum CE value of 2.9\% is observed for the short and intermediate pulse durations, compared with a value of 2.6\% obtained for the longest pulse duration. We note that the small observed changes in CE may lie within the systematic uncertainties of the experiment. The rather small changes in CE with pulse duration, as well as laser intensity, imply a near-linear scaling of in-band energy with laser energy. The in-band energy (Fig.\,\ref{ch5-fig:2}(i)) indeed almost doubles when doubling the laser pulse duration, and up to 3.8\,mJ of in-band energy are measured per pulse. The angular distributions of in-band EUV emission are all similar up to an angle of \SI{64}{\degree} from which point the 7.4-ns case shows increased emission. The anisotropy factor in this 7.4-ns case (see Fig.\,\ref{ch5-fig:2}(h)) roughly follows the one of the short and medium pulse duration cases up to a target diameter of 180\,\micron, after which it remains roughly constant at values around 0.6. This observed difference may be attributed to an increased EUV emission volume (which increases with pulse duration), partially extending beyond the liquid disk target, which enables radiating into the forward hemisphere. Another factor contributing to this difference may be a slight tilt of the disk target due to finite drift of the alignment of the laser to the droplet \cite{Reijers2018laser}.

The recorded spectra, shown in Fig.\,\ref{ch5-fig:2}(n), are remarkably similar when comparing the different laser pulse lengths. The only minor visible difference is the slightly wider main emission feature in the 7.4\,ns case. SP also remains virtually independent of laser pulse duration and follows the same trend as discussed as in Sec.\,\ref{subsec:2um-target-size}. For the longest pulse duration case, we note that the decrease in SP with increasing target size is no longer fully compensated by increases in the anisotropy factor and this is reflected in slightly lower overall CE values.

No major changes were thus observed in the scaling of emission characteristics with target size when changing laser pulse length, besides a levelling off of the anisotropy factor for the longest pulse length. In-band EUV emission is also here seen to follow linearly the input laser energy.


\section{EUV generation using 1- and 2-\texorpdfstring{$\boldsymbol{\upmu}$\lowercase{m}}{um} laser light}
\label{sec:1-vs-2um}
\begin{figure*}[tb]
		\centering
		\includegraphics[scale=1]{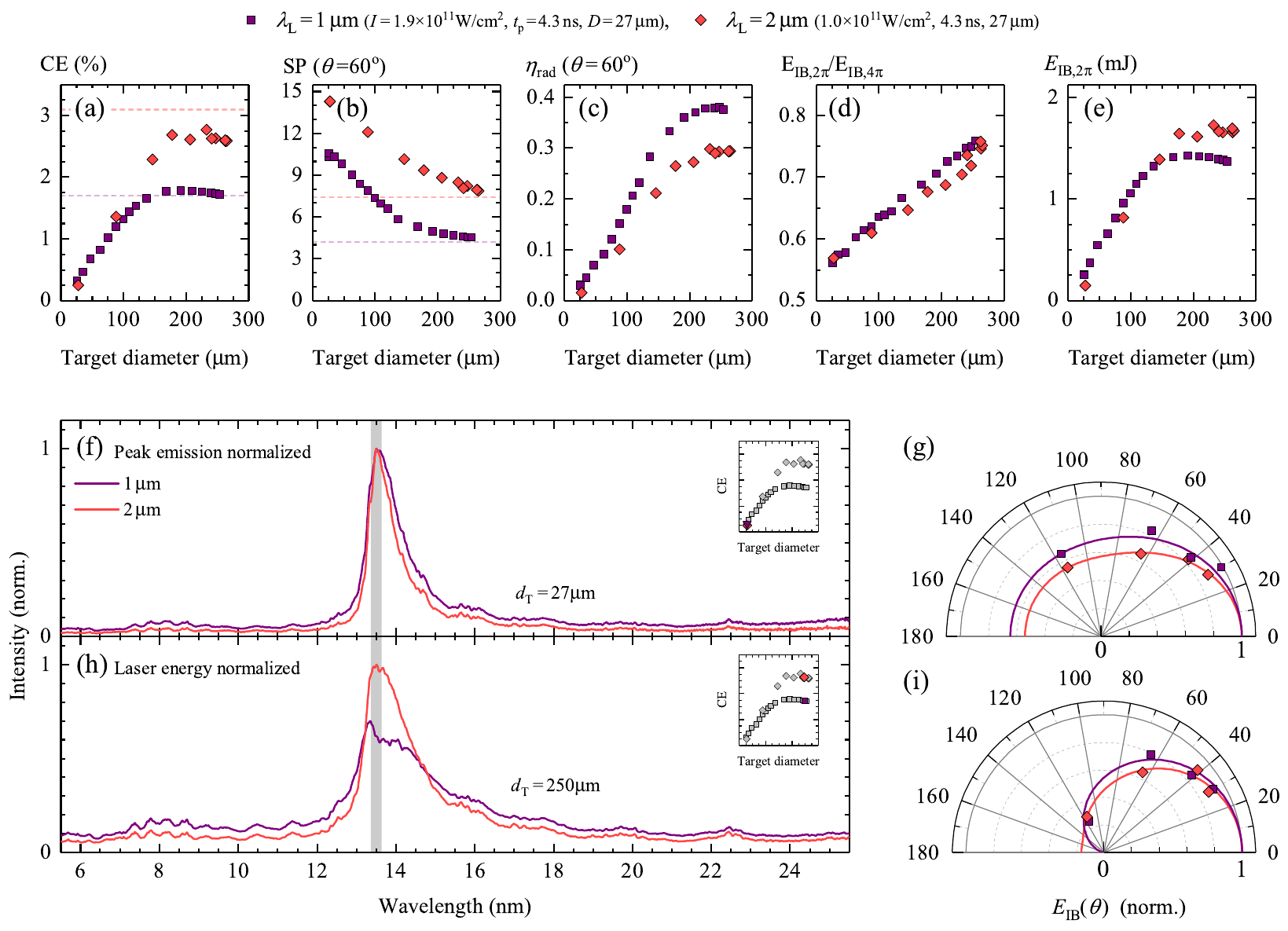}
		\caption{Comparison of results for drive laser beams of 1- and 2-\micron wavelength at intensities of 1.9 and \SI{1.0e11}{\intensity}, respectively. (a) CE, (b) SP, (c) $\eta_\mathrm{rad}$, (d) anisotropy factor $E_{\mathrm{IB},2\pi}/E_{\mathrm{IB},4\pi}$ and (e) $E_{\mathrm{IB},2\pi}$, as function of target diameter. The dashed lines in (a) and (b) indicate the CE and SP values from planar-solid tin targets, respectively \cite{Behnke2021}. The spectral emission of plasmas for a target diameter of (f) 27\,\micron (droplet target) and (h) 250\,\micron (disk target) are shown normalized to their respective peak intensities. The  1-\micron spectrum shown in (h) is normalized to laser energy with respect to the 2-\micron one.	The at \SI{0}{\degree} normalized angular dependencies of the in-band EUV emission, corresponding to the spectra shown in (f) and (h), are shown in (g) and (i), respectively. Solid lines depict fits of Eq.\,\eqref{eq:anisotropy-fit} to the data.}
	\label{ch5-fig:3}
\end{figure*}

In a separate, third set of experiments, plasma is produced using the 1-\micron laser beam at an intensity of \SI{1.9(4)e11}{\intensity}. This choice of laser intensity yields identical emission features in the 5--12\,nm region compared to the \SI[allow-number-unit-breaks]{1.0(2)e11}{\intensity} intensity at 2\,\micron. Equivalent spectral features ensure a similar charge state distribution of the plasma as each feature is charge-state specific. The factor of 1.9 difference in relative intensity is in agreement with the intensity ratio of 2.1(6) observed for droplet targets \cite{Schupp2020} and 2.4(7) for planar-solid targets \cite{Behnke2021}. 

As shown in Fig.\,\ref{ch5-fig:3}(a), CE is found to increase monotonically with increasing target size in the 2-\micron case until the laser beam reaches full overlap with the target, as described in Sec.\,\ref{sec:2um}. 
In line with the findings in Ref.\,\cite{sizyuk2020tuning} CE is virtually identical for both laser wavelength cases for small target sizes.
With increasing target diameter, the 2-\micron CE values exceed the ones of the 1-\micron case however significantly.
The 1-\micron CE is found to increase in a near-linear fashion until it plateaus at a value of close to 1.8\% above ${\sim}150$\,\micron target diameter. In the case of planar-solid targets (effectively infinitely extended disk targets), very similar CE values of 1.7\% were inferred from measurements performed at a representative angle of \SI{60}{\degree} for a comparable 4.8\,ns pulse duration and a slightly smaller circular 66-\micron-diameter (FWHM) spot\,\cite{Behnke2021}. 
In the 2-\micron case, CE values measured on planar-solid targets reach up to 3.1\%, only slightly higher than the here-observed values of up to 2.8\%.
	
The angular dependence of the in-band EUV emission for the droplet target (see Fig.\,\ref{ch5-fig:3}(g)) is very similar for both drive lasers and is consistent with previous studies where a 1-\micron laser was used to drive the LPP from a droplet target \cite{Yamaura2005,Giovannini2013,Schupp2019}. The similarities in the angular distributions is in line with a previous study by Chen \textit{et al.}\,\cite{chen2015angular} where only minor changes in emission anisotropy were found between 1- and 10-\micron-driven droplet plasmas. 
Similarly to Sec.\,\ref{sec:2um}, the in-band emission decreases more strongly with increasing angle, in particular for the angles $\gtrapprox$\SI{90}{\degree}, for the 250\,\micron target diameter than for the undeformed droplet target case (see Fig.\,\ref{ch5-fig:3}(i)).

For the droplet target the anisotropy factor attains a value of 0.57, identical for both laser wavelength cases (see Fig.\,\ref{ch5-fig:3}(d)). In both laser wavelength cases, the anisotropy factor increases linearly with increasing target diameter as discussed above in Sec.\,\ref{subsec:laserintensity}.  
Uncertainties related to the slightly worse fit of the anisotropy function to the 1-\micron disk-target data, which unexpectedly converges to 0 at 180$^\circ$, may lead to a minor systematic overestimation of the anisotropy factor. 

The in-band energy values displayed in Fig.\,\ref{ch5-fig:3}(e) trivially follow the scaling of CE. In the 1-\micron case, values of approximately 1.4\,mJ per pulse are achieved while reaching approximately 1.7\,mJ in the 2-\micron case. The relative difference in the in-band energy values is slightly smaller than for the CE values because of the lower laser energy used in the 2-\micron case (viz. 62 vs 80\,mJ pulse energy). Despite this smaller laser energy more in-band energy is produced by the 2-\micron pulses at target diameters above 160\,\micron demonstrating here the significant advantage of the 2x longer drive laser wavelength.

For undeformed droplet targets, the spectra for the 2-\micron case show a much narrower main emission feature and, relative to the 13.5-nm peak, less out-of-band radiation is emitted in the 2-\micron case (see Fig\,\ref{ch5-fig:3}(f)). These observations are consistent with earlier results\,\cite{Schupp2019,Schupp2020}. With increasing target diameter the differences in the spectral emission between the two drive laser wavelengths become even more pronounced. In the 2-\micron case, the main emission feature simply broadens with target diameter (see Fig.\,\ref{ch5-fig:2}(h)), in line with expectations from an associated increase in optical depth \cite{Schupp2019b}. In the 1-\micron case, the shape of the main emission feature changes drastically due to strong self-absorption, causing the emission maximum to shift to shorter wavelength. These absorption effects become even stronger for increased laser pulse duration, redistributing even more light into other energy channels (see Appendix\,\ref{sec:1-vs-2um_7.4ns}). This \q{missing} emission in the main feature in the 1-\micron case is qualitatively well explained by absorption of in-band radiation from a hotter plasma zone by a colder less emissive plasma zone \cite{Apruzese2002, Behnke2021}. 

SP is found to decrease monotonously with increasing target diameter for both wavelength cases (see Fig.\,\ref{ch5-fig:3}(b)). 
In the 1-\micron case, the SP for small target diameters is on the order of 10\%, and is observed to decrease more rapidly than its 2-\micron counterpart down to 4.5\% at a 250\,\micron diameter. 
This value is consistent with the marginally lower SP value of 4\% obtained from measurements on planar-solid tin targets which may be taken as the asymptotic value of SP towards infinite target size\,\cite{Behnke2021}. SP in the 2-\micron case levels off at a higher value of about 8\%, in part explaining the higher CE values observed using this drive laser wavelength.

Radiative efficiency, measured at an angle of \SI{60}{\degree}, follows a very similar trend in both wavelength cases (see panel (c)) with a larger amplitude in the 1-\micron case due to the significantly lower SP values in this laser wavelength case. 

No major differences between the anisotropy factors are observed. Comparing the two drive laser wavelength cases, the differences in CE are largely attributable to the decrease in SP.
This link between CE and SP further supports the findings of Behnke \textit{et al.}\,\cite{Behnke2021}, where experiments on solid tin targets demonstrated that plasmas driven by 1-\micron laser light exhibit strong EUV self-absorption which is absent in the 2-\micron spectra. This makes 2-\micron-driven plasmas the more efficient emitter of in-band EUV radiation. 
We note that conversion efficiencies of up to 3\% can in fact be achieved for the 1-\micron drive laser case for a homogeneous heating of undeformed droplet targets with several 10-ns-long, spatially flattop-shaped laser pulses \cite{Schupp2019}. Thus, on one hand, the current limitations to CE for the 1-\micron drive laser case for extended disk targets may be eased using alternate target and illumination designs. On the other hand, even larger CE values for the 2-\micron drive laser case may be obtained from optimally shaped targets homogeneously heated by a long, flattop laser pulse.

\section{Conclusions}
We have studied plasmas produced from laser pre-pulse preformed liquid tin disk targets with diameters ranging 30--300\,\micron using 1- and 2-\micron drive laser systems. For the 2-\micron driver, the conversion efficiency of laser energy to EUV radiation closely follows the fraction of the laser energy absorbed by the tin target and CE values of up to 3\% are obtained for the largest targets. 
Conversion efficiency (CE), spectral purity (SP), radiative efficiency ($\eta_\mathrm{rad}$), and spectral emission are found to be nearly independent of laser intensity and laser pulse duration in the here-studied parameter range. Consequently, a linear increase of in-band radiation towards the backward hemisphere with laser energy is observed when increasing either parameter and further scaling of in-band output per tin target with laser intensity and pulse duration may be possible at little to no cost regarding CE.

Direct comparison of the emission characteristics of 1- and 2-\micron-driven plasmas reveals significantly lower CE values for the 1-\micron driver under the current experimental conditions when using extended disk targets. The lower 1-\micron CE is explained by the particularly strong self-absorption of the emitted EUV radiation in the 1-\micron-driven plasma.
Further improvements in terms of CE may be obtainable by homogeneous heating of suitably shaped tin targets with longer laser pulses leading to full evaporation of the then truly \q{mass-limited} tin target. Such future studies should include research on the influence of laser pulse duration, intensity and laser wavelength on any possible fast ionic or liquid debris that may harm nearby optics elements in possible future industrial EUV light sources based on 2-\micron-laser-driven tin plasmas.

\begin{acknowledgments}
    This work has been carried out at the Advanced Research Center for Nanolithography (ARCNL), a public-private partnership of the University of Amsterdam (UvA), the Vrije Universiteit Amsterdam (VU), the Dutch Research Council (NWO) and the semiconductor equipment manufacturer ASML.
	The used transmission grating spectrometer has been developed in the Industrial Focus Group XUV Optics at University of Twente, and supported by the FOM Valorisation Prize 2011 awarded to F. Bijkerk and NanoNextNL Valorization Grant awarded to M. Bayraktar in 2015.
	This project has received funding from European Research Council (ERC) Starting Grant number 802648. This publication is part of the project New Light for Nanolithography (with project number 15697) of the research programme VIDI which is (partly) financed by the Dutch Research Council (NWO).
\end{acknowledgments}

\appendix

\section{EUV generation using longer pulses of 1- and 2-\texorpdfstring{$\boldsymbol{\upmu}$\lowercase{m}}{um} wavelength --- long pulse duration}
    \label{sec:1-vs-2um_7.4ns}
	
In addition to the comparison between 1- and 2-\micron-driven plasmas for the short-pulse case in Sec.\,\ref{sec:1-vs-2um}, more data is obtained using longer 1-\micron pulses that mimic the temporal profile of the 7.4\,ns case in Sec.\,\ref{subsec:pulseduration}.
The parameters of CE, SP, $\eta_\mathrm{rad}$, $E_\mathrm{IB,2\pi}/E_\mathrm{IB,4\pi}$ and $E_\mathrm{IB,2\pi}$ show similar trends compared to the short-pulse data in Sec.\,\ref{subsec:pulseduration}.

\begin{figure*}[h]
	\centering
	\includegraphics[scale=1]{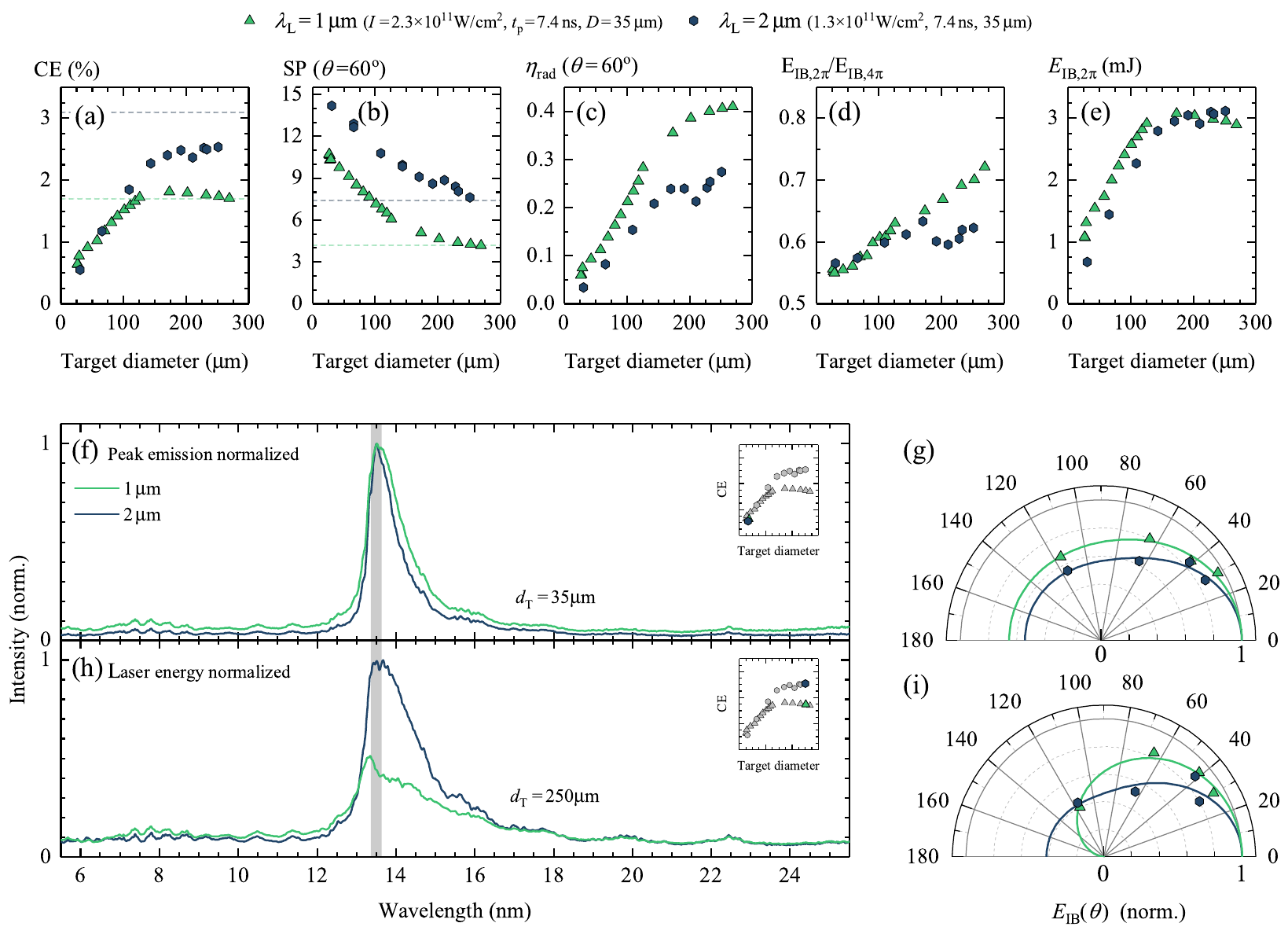}
	\caption{Comparison of results for drive laser beams of 1- and 2-\micron wavelength at intensities of 2.3 and \SI{1.3e11}{\intensity}, respectively, but for a longer pulse duration of 7.4\,ns. (a) CE, (b) SP, (c) $\eta_\mathrm{rad}$, (d) anisotropy factor $E_{\mathrm{IB},2\pi}/E_{\mathrm{IB},4\pi}$ and (e) $E_{\mathrm{IB},2\pi}$, as function of target diameter. The dashed lines in (a) and (b) indicate the CE and SP values from planar-solid tin targets, respectively \cite{Behnke2021}. The spectral emission of plasmas for a target diameter of (f) 35\,\micron (droplet target) and (h) 250\,\micron (disk target) are shown normalized to their respective peak intensities. The  1-\micron spectrum shown in (h) is normalized to laser energy, instead of its peak emission. The at \SI{0}{\degree} normalized angular dependencies of the in-band EUV emission, corresponding to the spectra shown in (f) and (h), are shown in (g) and (i), respectively. Solid lines depict fits of Eq.\,\eqref{eq:anisotropy-fit} to the data.}
	\label{ch5-fig:4}
\end{figure*}

The only significant change in comparison with the short-pulse data is observed in the spectral emission of the two drive-laser wavelength cases for disk-targets in Fig.\,\ref{ch5-fig:4}(h). Here, the spectral emission of plasmas produced with the 1-\micron laser beam show even stronger self-absorption of emission at and around 13.5\,nm. This stronger absorption manifests itself in slightly lower SP values. The slight deviation in the trend of $\eta_\mathrm{rad}$ above 160\,\micron target diameter in the 2-\micron case was already discussed in the main text.

In summary, good spectral performance of the plasma and high CE values are continued to be observed when increasing laser pulse duration of the 2-\micron driver. In contrast, increasing the pulse duration of the 1-\micron-wavelength driver, self-absorption effects increase significantly and lead to an even larger unwanted redistributing of in-band radiation into other energy channels than in the short-pulse case. 

\onecolumngrid

\twocolumngrid


\begin{thebibliography}{64}%
\makeatletter
\providecommand \@ifxundefined [1]{%
 \@ifx{#1\undefined}
}%
\providecommand \@ifnum [1]{%
 \ifnum #1\expandafter \@firstoftwo
 \else \expandafter \@secondoftwo
 \fi
}%
\providecommand \@ifx [1]{%
 \ifx #1\expandafter \@firstoftwo
 \else \expandafter \@secondoftwo
 \fi
}%
\providecommand \natexlab [1]{#1}%
\providecommand \enquote  [1]{``#1''}%
\providecommand \bibnamefont  [1]{#1}%
\providecommand \bibfnamefont [1]{#1}%
\providecommand \citenamefont [1]{#1}%
\providecommand \href@noop [0]{\@secondoftwo}%
\providecommand \href [0]{\begingroup \@sanitize@url \@href}%
\providecommand \@href[1]{\@@startlink{#1}\@@href}%
\providecommand \@@href[1]{\endgroup#1\@@endlink}%
\providecommand \@sanitize@url [0]{\catcode `\\12\catcode `\$12\catcode
  `\&12\catcode `\#12\catcode `\^12\catcode `\_12\catcode `\%12\relax}%
\providecommand \@@startlink[1]{}%
\providecommand \@@endlink[0]{}%
\providecommand \url  [0]{\begingroup\@sanitize@url \@url }%
\providecommand \@url [1]{\endgroup\@href {#1}{\urlprefix }}%
\providecommand \urlprefix  [0]{URL }%
\providecommand \Eprint [0]{\href }%
\providecommand \doibase [0]{https://doi.org/}%
\providecommand \selectlanguage [0]{\@gobble}%
\providecommand \bibinfo  [0]{\@secondoftwo}%
\providecommand \bibfield  [0]{\@secondoftwo}%
\providecommand \translation [1]{[#1]}%
\providecommand \BibitemOpen [0]{}%
\providecommand \bibitemStop [0]{}%
\providecommand \bibitemNoStop [0]{.\EOS\space}%
\providecommand \EOS [0]{\spacefactor3000\relax}%
\providecommand \BibitemShut  [1]{\csname bibitem#1\endcsname}%
\let\auto@bib@innerbib\@empty
\bibitem [{\citenamefont {Versolato}(2019)}]{versolato2019physics}%
  \BibitemOpen
  \bibfield  {author} {\bibinfo {author} {\bibfnamefont {O.~O.}\ \bibnamefont
  {Versolato}},\ }\bibfield  {title} {\bibinfo {title} {{Physics of
  laser-driven tin plasma sources of {EUV} radiation for nanolithography}},\
  }\href {https://doi.org/10.1088/1361-6595/ab3302} {\bibfield  {journal}
  {\bibinfo  {journal} {Plasma Sources Sci. Technol.}\ }\textbf {\bibinfo
  {volume} {28}},\ \bibinfo {pages} {083001} (\bibinfo {year}
  {2019})}\BibitemShut {NoStop}%
\bibitem [{\citenamefont {Purvis}\ \emph {et~al.}(2018)\citenamefont {Purvis},
  \citenamefont {Fomenkov}, \citenamefont {Schafgans}, \citenamefont {Vargas},
  \citenamefont {Rich}, \citenamefont {Tao}, \citenamefont {Rokitski},
  \citenamefont {Mulder}, \citenamefont {Buurman}, \citenamefont {Kats},
  \citenamefont {Stewart}, \citenamefont {LaForge}, \citenamefont {Rajyaguru},
  \citenamefont {Vaschenko}, \citenamefont {Ershov}, \citenamefont {Rafac},
  \citenamefont {Abraham}, \citenamefont {Brandt},\ and\ \citenamefont
  {Brown}}]{Purvis2018industrialization}%
  \BibitemOpen
  \bibfield  {author} {\bibinfo {author} {\bibfnamefont {M.}~\bibnamefont
  {Purvis}}, \bibinfo {author} {\bibfnamefont {I.~V.}\ \bibnamefont
  {Fomenkov}}, \bibinfo {author} {\bibfnamefont {A.~A.}\ \bibnamefont
  {Schafgans}}, \bibinfo {author} {\bibfnamefont {M.}~\bibnamefont {Vargas}},
  \bibinfo {author} {\bibfnamefont {S.}~\bibnamefont {Rich}}, \bibinfo {author}
  {\bibfnamefont {Y.}~\bibnamefont {Tao}}, \bibinfo {author} {\bibfnamefont
  {S.~I.}\ \bibnamefont {Rokitski}}, \bibinfo {author} {\bibfnamefont
  {M.}~\bibnamefont {Mulder}}, \bibinfo {author} {\bibfnamefont
  {E.}~\bibnamefont {Buurman}}, \bibinfo {author} {\bibfnamefont
  {M.}~\bibnamefont {Kats}}, \bibinfo {author} {\bibfnamefont {J.}~\bibnamefont
  {Stewart}}, \bibinfo {author} {\bibfnamefont {A.~D.}\ \bibnamefont
  {LaForge}}, \bibinfo {author} {\bibfnamefont {C.}~\bibnamefont {Rajyaguru}},
  \bibinfo {author} {\bibfnamefont {G.}~\bibnamefont {Vaschenko}}, \bibinfo
  {author} {\bibfnamefont {A.~I.}\ \bibnamefont {Ershov}}, \bibinfo {author}
  {\bibfnamefont {R.~J.}\ \bibnamefont {Rafac}}, \bibinfo {author}
  {\bibfnamefont {M.}~\bibnamefont {Abraham}}, \bibinfo {author} {\bibfnamefont
  {D.~C.}\ \bibnamefont {Brandt}},\ and\ \bibinfo {author} {\bibfnamefont
  {D.~J.}\ \bibnamefont {Brown}},\ }\bibfield  {title} {\bibinfo {title}
  {{Industrialization of a robust EUV source for high-volume manufacturing and
  power scaling beyond 250\,W}},\ }in\ \href
  {https://doi.org/10.1117/12.2305955} {\emph {\bibinfo {booktitle} {{EUV
  Lithography IX}}}},\ Vol.\ \bibinfo {volume} {10583}\ (\bibinfo
  {organization} {SPIE},\ \bibinfo {year} {2018})\ p.\ \bibinfo {pages}
  {1058327}\BibitemShut {NoStop}%
\bibitem [{\citenamefont {Moore}(2018)}]{Moore2018euv}%
  \BibitemOpen
  \bibfield  {author} {\bibinfo {author} {\bibfnamefont {S.~K.}\ \bibnamefont
  {Moore}},\ }\bibfield  {title} {\bibinfo {title} {{EUV lithography finally
  ready for fabs}},\ }\href {https://doi.org/10.1109/MSPEC.2018.8241736}
  {\bibfield  {journal} {\bibinfo  {journal} {IEEE Spectrum}\ }\textbf
  {\bibinfo {volume} {55}},\ \bibinfo {pages} {46} (\bibinfo {year}
  {2018})}\BibitemShut {NoStop}%
\bibitem [{\citenamefont {Schafgans}\ \emph {et~al.}(2015)\citenamefont
  {Schafgans}, \citenamefont {Brown}, \citenamefont {Fomenkov}, \citenamefont
  {Sandstrom}, \citenamefont {Ershov}, \citenamefont {Va\-schen\ ko},
  \citenamefont {Rafac}, \citenamefont {Purvis}, \citenamefont {Rokitski},
  \citenamefont {Tao}, \citenamefont {Riggs}, \citenamefont {Dunstan},
  \citenamefont {Graham}, \citenamefont {Farrar}, \citenamefont {Brandt},
  \citenamefont {Böwering}, \citenamefont {Pirati}, \citenamefont {Harned},
  \citenamefont {Wagner}, \citenamefont {Meiling},\ and\ \citenamefont
  {Kool}}]{Schafgans2015performance}%
  \BibitemOpen
  \bibfield  {author} {\bibinfo {author} {\bibfnamefont {A.~A.}\ \bibnamefont
  {Schafgans}}, \bibinfo {author} {\bibfnamefont {D.~J.}\ \bibnamefont
  {Brown}}, \bibinfo {author} {\bibfnamefont {I.~V.}\ \bibnamefont {Fomenkov}},
  \bibinfo {author} {\bibfnamefont {R.}~\bibnamefont {Sandstrom}}, \bibinfo
  {author} {\bibfnamefont {A.}~\bibnamefont {Ershov}}, \bibinfo {author}
  {\bibfnamefont {G.}~\bibnamefont {Va\-schen\ ko}}, \bibinfo {author}
  {\bibfnamefont {R.}~\bibnamefont {Rafac}}, \bibinfo {author} {\bibfnamefont
  {M.}~\bibnamefont {Purvis}}, \bibinfo {author} {\bibfnamefont
  {S.}~\bibnamefont {Rokitski}}, \bibinfo {author} {\bibfnamefont
  {Y.}~\bibnamefont {Tao}}, \bibinfo {author} {\bibfnamefont {D.~J.}\
  \bibnamefont {Riggs}}, \bibinfo {author} {\bibfnamefont {W.~J.}\ \bibnamefont
  {Dunstan}}, \bibinfo {author} {\bibfnamefont {M.}~\bibnamefont {Graham}},
  \bibinfo {author} {\bibfnamefont {N.~R.}\ \bibnamefont {Farrar}}, \bibinfo
  {author} {\bibfnamefont {D.~C.}\ \bibnamefont {Brandt}}, \bibinfo {author}
  {\bibfnamefont {N.}~\bibnamefont {Böwering}}, \bibinfo {author}
  {\bibfnamefont {A.}~\bibnamefont {Pirati}}, \bibinfo {author} {\bibfnamefont
  {N.}~\bibnamefont {Harned}}, \bibinfo {author} {\bibfnamefont
  {C.}~\bibnamefont {Wagner}}, \bibinfo {author} {\bibfnamefont
  {H.}~\bibnamefont {Meiling}},\ and\ \bibinfo {author} {\bibfnamefont
  {R.}~\bibnamefont {Kool}},\ }\bibfield  {title} {\bibinfo {title}
  {{Performance optimization of MOPA pre-pulse LPP light source}},\ }in\ \href
  {https://doi.org/10.1117/12.2087421} {\emph {\bibinfo {booktitle} {{EUV
  Lithography VI}}}},\ Vol.\ \bibinfo {volume} {9422}\ (\bibinfo {organization}
  {SPIE},\ \bibinfo {year} {2015})\ p.\ \bibinfo {pages} {94220B}\BibitemShut
  {NoStop}%
\bibitem [{\citenamefont {Mizoguchi}\ \emph {et~al.}(2018)\citenamefont
  {Mizoguchi}, \citenamefont {Nakarai}, \citenamefont {Abe}, \citenamefont
  {Nowak}, \citenamefont {Kawasuji}, \citenamefont {Tanaka}, \citenamefont
  {Watanabe}, \citenamefont {Hori}, \citenamefont {Kodama}, \citenamefont
  {Shiraishi}, \citenamefont {Yanagida}, \citenamefont {Soumagne},
  \citenamefont {Yamada}, \citenamefont {Yamazaki},\ and\ \citenamefont
  {Saitou}}]{mizoguchi2018high}%
  \BibitemOpen
  \bibfield  {author} {\bibinfo {author} {\bibfnamefont {H.}~\bibnamefont
  {Mizoguchi}}, \bibinfo {author} {\bibfnamefont {H.}~\bibnamefont {Nakarai}},
  \bibinfo {author} {\bibfnamefont {T.}~\bibnamefont {Abe}}, \bibinfo {author}
  {\bibfnamefont {K.~M.}\ \bibnamefont {Nowak}}, \bibinfo {author}
  {\bibfnamefont {Y.}~\bibnamefont {Kawasuji}}, \bibinfo {author}
  {\bibfnamefont {H.}~\bibnamefont {Tanaka}}, \bibinfo {author} {\bibfnamefont
  {Y.}~\bibnamefont {Watanabe}}, \bibinfo {author} {\bibfnamefont
  {T.}~\bibnamefont {Hori}}, \bibinfo {author} {\bibfnamefont {T.}~\bibnamefont
  {Kodama}}, \bibinfo {author} {\bibfnamefont {Y.}~\bibnamefont {Shiraishi}},
  \bibinfo {author} {\bibfnamefont {T.}~\bibnamefont {Yanagida}}, \bibinfo
  {author} {\bibfnamefont {G.}~\bibnamefont {Soumagne}}, \bibinfo {author}
  {\bibfnamefont {T.}~\bibnamefont {Yamada}}, \bibinfo {author} {\bibfnamefont
  {T.}~\bibnamefont {Yamazaki}},\ and\ \bibinfo {author} {\bibfnamefont
  {T.}~\bibnamefont {Saitou}},\ }\bibfield  {title} {\bibinfo {title} {{High
  power LPP-EUV source with long collector mirror lifetime for high volume
  semiconductor manufacturing}},\ }in\ \href
  {https://doi.org/https://doi.org/10.1109/CSTIC.2018.8369210} {\emph {\bibinfo
  {booktitle} {{Proc. CSTIC 2018}}}}\ (\bibinfo {organization} {IEEE},\
  \bibinfo {year} {2018})\ pp.\ \bibinfo {pages} {1--7}\BibitemShut {NoStop}%
\bibitem [{\citenamefont {O'Sullivan}\ \emph {et~al.}(2015)\citenamefont
  {O'Sullivan}, \citenamefont {Li}, \citenamefont {D'Arcy}, \citenamefont
  {Dunne}, \citenamefont {Hayden}, \citenamefont {Kilbane}, \citenamefont
  {McCormack}, \citenamefont {Ohashi}, \citenamefont {O'Reilly}, \citenamefont
  {Sheridan}, \citenamefont {Sokell}, \citenamefont {Suzuki},\ and\
  \citenamefont {Higashiguchi}}]{OSullivan2015}%
  \BibitemOpen
  \bibfield  {author} {\bibinfo {author} {\bibfnamefont {G.}~\bibnamefont
  {O'Sullivan}}, \bibinfo {author} {\bibfnamefont {B.}~\bibnamefont {Li}},
  \bibinfo {author} {\bibfnamefont {R.}~\bibnamefont {D'Arcy}}, \bibinfo
  {author} {\bibfnamefont {P.}~\bibnamefont {Dunne}}, \bibinfo {author}
  {\bibfnamefont {P.}~\bibnamefont {Hayden}}, \bibinfo {author} {\bibfnamefont
  {D.}~\bibnamefont {Kilbane}}, \bibinfo {author} {\bibfnamefont
  {T.}~\bibnamefont {McCormack}}, \bibinfo {author} {\bibfnamefont
  {H.}~\bibnamefont {Ohashi}}, \bibinfo {author} {\bibfnamefont
  {F.}~\bibnamefont {O'Reilly}}, \bibinfo {author} {\bibfnamefont
  {P.}~\bibnamefont {Sheridan}}, \bibinfo {author} {\bibfnamefont
  {E.}~\bibnamefont {Sokell}}, \bibinfo {author} {\bibfnamefont
  {C.}~\bibnamefont {Suzuki}},\ and\ \bibinfo {author} {\bibfnamefont
  {T.}~\bibnamefont {Higashiguchi}},\ }\bibfield  {title} {\bibinfo {title}
  {{Spectroscopy of highly charged ions and its relevance to EUV and soft x-ray
  source development}},\ }\href
  {https://doi.org/10.1088/0953-4075/48/14/144025} {\bibfield  {journal}
  {\bibinfo  {journal} {J. Phys. B: At. Mol. Opt. Phys.}\ }\textbf {\bibinfo
  {volume} {48}},\ \bibinfo {pages} {144025} (\bibinfo {year}
  {2015})}\BibitemShut {NoStop}%
\bibitem [{\citenamefont {{A. Yu. Vinokhodov}}\ \emph
  {et~al.}(2016)\citenamefont {{A. Yu. Vinokhodov}}, \citenamefont
  {Krivokorytov}, \citenamefont {{Yu. V. Sidelnikov}}, \citenamefont
  {Krivtsun}, \citenamefont {Medvedev},\ and\ \citenamefont
  {Koshelev}}]{Vinokhodov2016droplet}%
  \BibitemOpen
  \bibfield  {author} {\bibinfo {author} {\bibnamefont {{A. Yu. Vinokhodov}}},
  \bibinfo {author} {\bibfnamefont {M.~S.}\ \bibnamefont {Krivokorytov}},
  \bibinfo {author} {\bibnamefont {{Yu. V. Sidelnikov}}}, \bibinfo {author}
  {\bibfnamefont {V.~M.}\ \bibnamefont {Krivtsun}}, \bibinfo {author}
  {\bibfnamefont {V.~V.}\ \bibnamefont {Medvedev}},\ and\ \bibinfo {author}
  {\bibfnamefont {K.~N.}\ \bibnamefont {Koshelev}},\ }\bibfield  {title}
  {\bibinfo {title} {{Droplet-based, high-brightness extreme ultraviolet laser
  plasma source for metrology}},\ }\href {https://doi.org/10.1063/1.4966930}
  {\bibfield  {journal} {\bibinfo  {journal} {J. Appl. Phys.}\ }\textbf
  {\bibinfo {volume} {120}},\ \bibinfo {pages} {163304} (\bibinfo {year}
  {2016})}\BibitemShut {NoStop}%
\bibitem [{\citenamefont {Harilal}\ \emph {et~al.}(2011)\citenamefont
  {Harilal}, \citenamefont {Sizyuk}, \citenamefont {Hassanein}, \citenamefont
  {Campos}, \citenamefont {Hough},\ and\ \citenamefont
  {Sizyuk}}]{Harilal2011effect}%
  \BibitemOpen
  \bibfield  {author} {\bibinfo {author} {\bibfnamefont {S.~S.}\ \bibnamefont
  {Harilal}}, \bibinfo {author} {\bibfnamefont {T.}~\bibnamefont {Sizyuk}},
  \bibinfo {author} {\bibfnamefont {A.}~\bibnamefont {Hassanein}}, \bibinfo
  {author} {\bibfnamefont {D.}~\bibnamefont {Campos}}, \bibinfo {author}
  {\bibfnamefont {P.}~\bibnamefont {Hough}},\ and\ \bibinfo {author}
  {\bibfnamefont {V.}~\bibnamefont {Sizyuk}},\ }\bibfield  {title} {\bibinfo
  {title} {{The effect of excitation wavelength on dynamics of laser-produced
  tin plasma}},\ }\href {https://doi.org/10.1063/1.3562143} {\bibfield
  {journal} {\bibinfo  {journal} {J. Appl. Phys.}\ }\textbf {\bibinfo {volume}
  {109}},\ \bibinfo {pages} {063306} (\bibinfo {year} {2011})}\BibitemShut
  {NoStop}%
\bibitem [{\citenamefont {Giovannini}\ and\ \citenamefont
  {Abhari}(2014)}]{Giovannini2014}%
  \BibitemOpen
  \bibfield  {author} {\bibinfo {author} {\bibfnamefont {A.~Z.}\ \bibnamefont
  {Giovannini}}\ and\ \bibinfo {author} {\bibfnamefont {R.~S.}\ \bibnamefont
  {Abhari}},\ }\bibfield  {title} {\bibinfo {title} {{Effects of the dynamics
  of droplet-based laser-produced plasma on angular extreme ultraviolet
  emission profile}},\ }\href {https://doi.org/10.1063/1.4878506} {\bibfield
  {journal} {\bibinfo  {journal} {Appl. Phys. Lett.}\ }\textbf {\bibinfo
  {volume} {104}},\ \bibinfo {pages} {194104} (\bibinfo {year}
  {2014})}\BibitemShut {NoStop}%
\bibitem [{\citenamefont {Banine}\ \emph {et~al.}(2011)\citenamefont {Banine},
  \citenamefont {Koshelev},\ and\ \citenamefont {Swinkels}}]{Banine2011}%
  \BibitemOpen
  \bibfield  {author} {\bibinfo {author} {\bibfnamefont {V.~Y.}\ \bibnamefont
  {Banine}}, \bibinfo {author} {\bibfnamefont {K.~N.}\ \bibnamefont
  {Koshelev}},\ and\ \bibinfo {author} {\bibfnamefont {G.~H. P.~M.}\
  \bibnamefont {Swinkels}},\ }\bibfield  {title} {\bibinfo {title} {{Physical
  processes in EUV sources for microlithography}},\ }\href
  {https://doi.org/10.1088/0022-3727/44/25/253001} {\bibfield  {journal}
  {\bibinfo  {journal} {J. Phys. D: Appl. Phys.}\ }\textbf {\bibinfo {volume}
  {44}},\ \bibinfo {pages} {253001} (\bibinfo {year} {2011})}\BibitemShut
  {NoStop}%
\bibitem [{\citenamefont {Benschop}\ \emph {et~al.}(2008)\citenamefont
  {Benschop}, \citenamefont {Banine}, \citenamefont {Lok},\ and\ \citenamefont
  {Loopstra}}]{Benschop2008}%
  \BibitemOpen
  \bibfield  {author} {\bibinfo {author} {\bibfnamefont {J.}~\bibnamefont
  {Benschop}}, \bibinfo {author} {\bibfnamefont {V.}~\bibnamefont {Banine}},
  \bibinfo {author} {\bibfnamefont {S.}~\bibnamefont {Lok}},\ and\ \bibinfo
  {author} {\bibfnamefont {E.}~\bibnamefont {Loopstra}},\ }\bibfield  {title}
  {\bibinfo {title} {{Extreme ultraviolet lithography: Status and prospects}},\
  }\href {https://doi.org/10.1116/1.3010737} {\bibfield  {journal} {\bibinfo
  {journal} {J. Vac. Sci. Technol. B}\ }\textbf {\bibinfo {volume} {26}},\
  \bibinfo {pages} {2204} (\bibinfo {year} {2008})}\BibitemShut {NoStop}%
\bibitem [{\citenamefont {Azarov}\ and\ \citenamefont
  {Joshi}(1993)}]{Azarov1993}%
  \BibitemOpen
  \bibfield  {author} {\bibinfo {author} {\bibfnamefont {V.~I.}\ \bibnamefont
  {Azarov}}\ and\ \bibinfo {author} {\bibfnamefont {Y.~N.}\ \bibnamefont
  {Joshi}},\ }\bibfield  {title} {\bibinfo {title} {{Analysis of the
  $4d^7$--$4d^6 5p$ transition array of the eighth spectrum of tin: Sn VIII}},\
  }\href {https://doi.org/10.1088/0953-4075/26/20/011} {\bibfield  {journal}
  {\bibinfo  {journal} {J. Phys. B: At. Mol. Opt. Phys.}\ }\textbf {\bibinfo
  {volume} {26}},\ \bibinfo {pages} {3495} (\bibinfo {year}
  {1993})}\BibitemShut {NoStop}%
\bibitem [{\citenamefont {Churilov}\ and\ \citenamefont
  {Ryabtsev}(2006{\natexlab{a}})}]{Churilov2006SnIX--SnXII}%
  \BibitemOpen
  \bibfield  {author} {\bibinfo {author} {\bibfnamefont {S.~S.}\ \bibnamefont
  {Churilov}}\ and\ \bibinfo {author} {\bibfnamefont {A.~N.}\ \bibnamefont
  {Ryabtsev}},\ }\bibfield  {title} {\bibinfo {title} {{Analyses of the Sn
  IX--Sn XII spectra in the EUV region}},\ }\href
  {https://doi.org/10.1088/0031-8949/73/6/014} {\bibfield  {journal} {\bibinfo
  {journal} {Phys. Scr.}\ }\textbf {\bibinfo {volume} {73}},\ \bibinfo {pages}
  {614} (\bibinfo {year} {2006}{\natexlab{a}})}\BibitemShut {NoStop}%
\bibitem [{\citenamefont {Churilov}\ and\ \citenamefont
  {Ryabtsev}(2006{\natexlab{b}})}]{Churilov2006SnVIII}%
  \BibitemOpen
  \bibfield  {author} {\bibinfo {author} {\bibfnamefont {S.~S.}\ \bibnamefont
  {Churilov}}\ and\ \bibinfo {author} {\bibfnamefont {A.~N.}\ \bibnamefont
  {Ryabtsev}},\ }\bibfield  {title} {\bibinfo {title} {{Analysis of the 4$p^6
  4d^7$--(4$p^6 4d^6 4f$ + $4p^5 4d^8$) transitions in the Sn VIII spectrum}},\
  }\href {https://doi.org/10.1134/S0030400X06050043} {\bibfield  {journal}
  {\bibinfo  {journal} {Opt. Spectrosc.}\ }\textbf {\bibinfo {volume} {100}},\
  \bibinfo {pages} {660} (\bibinfo {year} {2006}{\natexlab{b}})}\BibitemShut
  {NoStop}%
\bibitem [{\citenamefont {Churilov}\ and\ \citenamefont
  {Ryabtsev}(2006{\natexlab{c}})}]{Churilov2006SnXIII--XV}%
  \BibitemOpen
  \bibfield  {author} {\bibinfo {author} {\bibfnamefont {S.~S.}\ \bibnamefont
  {Churilov}}\ and\ \bibinfo {author} {\bibfnamefont {A.~N.}\ \bibnamefont
  {Ryabtsev}},\ }\bibfield  {title} {\bibinfo {title} {{Analysis of the spectra
  of In XII--XIV and Sn XIII--XV in the far-VUV region}},\ }\href
  {https://doi.org/10.1134/S0030400X06080017} {\bibfield  {journal} {\bibinfo
  {journal} {Opt. Spectrosc.}\ }\textbf {\bibinfo {volume} {101}},\ \bibinfo
  {pages} {169} (\bibinfo {year} {2006}{\natexlab{c}})}\BibitemShut {NoStop}%
\bibitem [{\citenamefont {Ryabtsev}\ \emph {et~al.}(2008)\citenamefont
  {Ryabtsev}, \citenamefont {{\'E. Ya. Kononov}},\ and\ \citenamefont
  {Churilov}}]{Ryabtsev2008SnXIV}%
  \BibitemOpen
  \bibfield  {author} {\bibinfo {author} {\bibfnamefont {A.~N.}\ \bibnamefont
  {Ryabtsev}}, \bibinfo {author} {\bibnamefont {{\'E. Ya. Kononov}}},\ and\
  \bibinfo {author} {\bibfnamefont {S.~S.}\ \bibnamefont {Churilov}},\
  }\bibfield  {title} {\bibinfo {title} {{Spectra of rubidium-like Pd X--Sn XIV
  ions}},\ }\href {https://doi.org/10.1134/S0030400X08120060} {\bibfield
  {journal} {\bibinfo  {journal} {Opt. Spectrosc.}\ }\textbf {\bibinfo {volume}
  {105}},\ \bibinfo {pages} {844} (\bibinfo {year} {2008})}\BibitemShut
  {NoStop}%
\bibitem [{\citenamefont {{I. Yu. Tolstikhina}}\ \emph
  {et~al.}(2006)\citenamefont {{I. Yu. Tolstikhina}}, \citenamefont {Churilov},
  \citenamefont {Ryabtsev},\ and\ \citenamefont
  {Koshelev}}]{Tolstikhina2006ATOMICDATA}%
  \BibitemOpen
  \bibfield  {author} {\bibinfo {author} {\bibnamefont {{I. Yu. Tolstikhina}}},
  \bibinfo {author} {\bibfnamefont {S.~S.}\ \bibnamefont {Churilov}}, \bibinfo
  {author} {\bibfnamefont {A.~N.}\ \bibnamefont {Ryabtsev}},\ and\ \bibinfo
  {author} {\bibfnamefont {K.~N.}\ \bibnamefont {Koshelev}},\ }\bibfield
  {title} {\bibinfo {title} {Atomic tin data},\ }in\ \href@noop {} {\emph
  {\bibinfo {booktitle} {{EUV sources for lithography}}}},\ \bibinfo {editor}
  {edited by\ \bibinfo {editor} {\bibfnamefont {V.}~\bibnamefont {Bakshi}}}\
  (\bibinfo  {publisher} {SPIE press Bellingham, Washington},\ \bibinfo {year}
  {2006})\ Chap.~\bibinfo {chapter} {4}, pp.\ \bibinfo {pages}
  {113--148}\BibitemShut {NoStop}%
\bibitem [{\citenamefont {D'Arcy}\ \emph {et~al.}(2009)\citenamefont {D'Arcy},
  \citenamefont {Ohashi}, \citenamefont {Suda}, \citenamefont {Tanuma},
  \citenamefont {Fujioka}, \citenamefont {Nishimura}, \citenamefont
  {Nishihara}, \citenamefont {Suzuki}, \citenamefont {Kato}, \citenamefont
  {Koike}, \citenamefont {White},\ and\ \citenamefont
  {O'Sullivan}}]{DArcy2009a}%
  \BibitemOpen
  \bibfield  {author} {\bibinfo {author} {\bibfnamefont {R.}~\bibnamefont
  {D'Arcy}}, \bibinfo {author} {\bibfnamefont {H.}~\bibnamefont {Ohashi}},
  \bibinfo {author} {\bibfnamefont {S.}~\bibnamefont {Suda}}, \bibinfo {author}
  {\bibfnamefont {H.}~\bibnamefont {Tanuma}}, \bibinfo {author} {\bibfnamefont
  {S.}~\bibnamefont {Fujioka}}, \bibinfo {author} {\bibfnamefont
  {H.}~\bibnamefont {Nishimura}}, \bibinfo {author} {\bibfnamefont
  {K.}~\bibnamefont {Nishihara}}, \bibinfo {author} {\bibfnamefont
  {C.}~\bibnamefont {Suzuki}}, \bibinfo {author} {\bibfnamefont
  {T.}~\bibnamefont {Kato}}, \bibinfo {author} {\bibfnamefont {F.}~\bibnamefont
  {Koike}}, \bibinfo {author} {\bibfnamefont {J.}~\bibnamefont {White}},\ and\
  \bibinfo {author} {\bibfnamefont {G.}~\bibnamefont {O'Sullivan}},\ }\bibfield
   {title} {\bibinfo {title} {{Transitions and the effects of configuration
  interaction in the spectra of Sn XV–Sn XVIII}},\ }\href
  {https://doi.org/10.1103/PhysRevA.79.042509} {\bibfield  {journal} {\bibinfo
  {journal} {Phys. Rev. A}\ }\textbf {\bibinfo {volume} {79}},\ \bibinfo
  {pages} {042509} (\bibinfo {year} {2009})}\BibitemShut {NoStop}%
\bibitem [{\citenamefont {Ohashi}\ \emph {et~al.}(2010)\citenamefont {Ohashi},
  \citenamefont {Suda}, \citenamefont {Tanuma}, \citenamefont {Fujioka},
  \citenamefont {Nishimura}, \citenamefont {Sasaki},\ and\ \citenamefont
  {Nishihara}}]{Ohashi2010}%
  \BibitemOpen
  \bibfield  {author} {\bibinfo {author} {\bibfnamefont {H.}~\bibnamefont
  {Ohashi}}, \bibinfo {author} {\bibfnamefont {S.}~\bibnamefont {Suda}},
  \bibinfo {author} {\bibfnamefont {H.}~\bibnamefont {Tanuma}}, \bibinfo
  {author} {\bibfnamefont {S.}~\bibnamefont {Fujioka}}, \bibinfo {author}
  {\bibfnamefont {H.}~\bibnamefont {Nishimura}}, \bibinfo {author}
  {\bibfnamefont {A.}~\bibnamefont {Sasaki}},\ and\ \bibinfo {author}
  {\bibfnamefont {K.}~\bibnamefont {Nishihara}},\ }\bibfield  {title} {\bibinfo
  {title} {{EUV emission spectra in collisions of multiply charged Sn ions with
  He and Xe}},\ }\href {https://doi.org/10.1088/0953-4075/43/6/065204}
  {\bibfield  {journal} {\bibinfo  {journal} {J. Phys. B: At. Mol. Opt. Phys.}\
  }\textbf {\bibinfo {volume} {43}},\ \bibinfo {pages} {065204} (\bibinfo
  {year} {2010})}\BibitemShut {NoStop}%
\bibitem [{\citenamefont {Colgan}\ \emph {et~al.}(2017)\citenamefont {Colgan},
  \citenamefont {Kilcrease}, \citenamefont {Abdallah}, \citenamefont
  {Sherrill}, \citenamefont {Fontes}, \citenamefont {Hakel},\ and\
  \citenamefont {Armstrong}}]{Colgan2017}%
  \BibitemOpen
  \bibfield  {author} {\bibinfo {author} {\bibfnamefont {J.}~\bibnamefont
  {Colgan}}, \bibinfo {author} {\bibfnamefont {D.~P.}\ \bibnamefont
  {Kilcrease}}, \bibinfo {author} {\bibfnamefont {J.}~\bibnamefont {Abdallah}},
  \bibinfo {author} {\bibfnamefont {M.~E.}\ \bibnamefont {Sherrill}}, \bibinfo
  {author} {\bibfnamefont {C.~J.}\ \bibnamefont {Fontes}}, \bibinfo {author}
  {\bibfnamefont {P.}~\bibnamefont {Hakel}},\ and\ \bibinfo {author}
  {\bibfnamefont {G.~S.~J.}\ \bibnamefont {Armstrong}},\ }\bibfield  {title}
  {\bibinfo {title} {{Atomic structure considerations for the low-temperature
  opacity of Sn}},\ }\href {https://doi.org/10.1016/j.hedp.2017.03.009}
  {\bibfield  {journal} {\bibinfo  {journal} {High Energy Density Phys.}\
  }\textbf {\bibinfo {volume} {23}},\ \bibinfo {pages} {133 } (\bibinfo {year}
  {2017})}\BibitemShut {NoStop}%
\bibitem [{\citenamefont {Torretti}\ \emph {et~al.}(2017)\citenamefont
  {Torretti}, \citenamefont {Windberger}, \citenamefont {Ryabtsev},
  \citenamefont {Dobrodey}, \citenamefont {Bekker}, \citenamefont {Ubachs},
  \citenamefont {Hoekstra}, \citenamefont {Kahl}, \citenamefont {Berengut},
  \citenamefont {Crespo L\'opez-Urrutia},\ and\ \citenamefont
  {Versolato}}]{Torretti2017}%
  \BibitemOpen
  \bibfield  {author} {\bibinfo {author} {\bibfnamefont {F.}~\bibnamefont
  {Torretti}}, \bibinfo {author} {\bibfnamefont {A.}~\bibnamefont
  {Windberger}}, \bibinfo {author} {\bibfnamefont {A.}~\bibnamefont
  {Ryabtsev}}, \bibinfo {author} {\bibfnamefont {S.}~\bibnamefont {Dobrodey}},
  \bibinfo {author} {\bibfnamefont {H.}~\bibnamefont {Bekker}}, \bibinfo
  {author} {\bibfnamefont {W.}~\bibnamefont {Ubachs}}, \bibinfo {author}
  {\bibfnamefont {R.}~\bibnamefont {Hoekstra}}, \bibinfo {author}
  {\bibfnamefont {E.~V.}\ \bibnamefont {Kahl}}, \bibinfo {author}
  {\bibfnamefont {J.~C.}\ \bibnamefont {Berengut}}, \bibinfo {author}
  {\bibfnamefont {J.~R.}\ \bibnamefont {Crespo L\'opez-Urrutia}},\ and\
  \bibinfo {author} {\bibfnamefont {O.~O.}\ \bibnamefont {Versolato}},\
  }\bibfield  {title} {\bibinfo {title} {{Optical spectroscopy of complex
  open-$4d$-shell ions Sn$^{7+}$--Sn$^{10+}$}},\ }\href
  {https://doi.org/10.1103/PhysRevA.95.042503} {\bibfield  {journal} {\bibinfo
  {journal} {Phys. Rev. A}\ }\textbf {\bibinfo {volume} {95}},\ \bibinfo
  {pages} {042503} (\bibinfo {year} {2017})}\BibitemShut {NoStop}%
\bibitem [{\citenamefont {Scheers}\ \emph {et~al.}(2020)\citenamefont
  {Scheers}, \citenamefont {Shah}, \citenamefont {Ryabtsev}, \citenamefont
  {Bekker}, \citenamefont {Torretti}, \citenamefont {Sheil}, \citenamefont
  {Czapski}, \citenamefont {Berengut}, \citenamefont {Ubachs}, \citenamefont
  {Crespo L\'opez-Urrutia}, \citenamefont {Hoekstra},\ and\ \citenamefont
  {Versolato}}]{Scheers2020}%
  \BibitemOpen
  \bibfield  {author} {\bibinfo {author} {\bibfnamefont {J.}~\bibnamefont
  {Scheers}}, \bibinfo {author} {\bibfnamefont {C.}~\bibnamefont {Shah}},
  \bibinfo {author} {\bibfnamefont {A.}~\bibnamefont {Ryabtsev}}, \bibinfo
  {author} {\bibfnamefont {H.}~\bibnamefont {Bekker}}, \bibinfo {author}
  {\bibfnamefont {F.}~\bibnamefont {Torretti}}, \bibinfo {author}
  {\bibfnamefont {J.}~\bibnamefont {Sheil}}, \bibinfo {author} {\bibfnamefont
  {D.~A.}\ \bibnamefont {Czapski}}, \bibinfo {author} {\bibfnamefont {J.~C.}\
  \bibnamefont {Berengut}}, \bibinfo {author} {\bibfnamefont {W.}~\bibnamefont
  {Ubachs}}, \bibinfo {author} {\bibfnamefont {J.~R.}\ \bibnamefont {Crespo
  L\'opez-Urrutia}}, \bibinfo {author} {\bibfnamefont {R.}~\bibnamefont
  {Hoekstra}},\ and\ \bibinfo {author} {\bibfnamefont {O.~O.}\ \bibnamefont
  {Versolato}},\ }\bibfield  {title} {\bibinfo {title} {{EUV spectroscopy of
  highly charged ${\mathrm{Sn}}^{13+}$--$\mathrm{Sn}^{15+}$ ions in an
  electron-beam ion trap}},\ }\href
  {https://doi.org/10.1103/PhysRevA.101.062511} {\bibfield  {journal} {\bibinfo
   {journal} {Phys. Rev. A}\ }\textbf {\bibinfo {volume} {101}},\ \bibinfo
  {pages} {062511} (\bibinfo {year} {2020})}\BibitemShut {NoStop}%
\bibitem [{\citenamefont {Bajt}\ \emph {et~al.}(2002)\citenamefont {Bajt},
  \citenamefont {Alameda}, \citenamefont {Barbee}, \citenamefont {Clift},
  \citenamefont {Folta}, \citenamefont {Kaufmann},\ and\ \citenamefont
  {Spiller}}]{Bajt2002}%
  \BibitemOpen
  \bibfield  {author} {\bibinfo {author} {\bibfnamefont {S.}~\bibnamefont
  {Bajt}}, \bibinfo {author} {\bibfnamefont {J.~B.}\ \bibnamefont {Alameda}},
  \bibinfo {author} {\bibfnamefont {T.~W.}\ \bibnamefont {Barbee},
  \bibfnamefont {Jr.}}, \bibinfo {author} {\bibfnamefont {W.~M.}\ \bibnamefont
  {Clift}}, \bibinfo {author} {\bibfnamefont {J.~A.}\ \bibnamefont {Folta}},
  \bibinfo {author} {\bibfnamefont {B.~B.}\ \bibnamefont {Kaufmann}},\ and\
  \bibinfo {author} {\bibfnamefont {E.~A.}\ \bibnamefont {Spiller}},\
  }\bibfield  {title} {\bibinfo {title} {{Improved reflectance and stability of
  Mo/Si multilayers}},\ }\href {https://doi.org/10.1117/1.1489426} {\bibfield
  {journal} {\bibinfo  {journal} {Opt. Eng.}\ }\textbf {\bibinfo {volume}
  {41}},\ \bibinfo {pages} {1797} (\bibinfo {year} {2002})}\BibitemShut
  {NoStop}%
\bibitem [{\citenamefont {Huang}\ \emph {et~al.}(2017)\citenamefont {Huang},
  \citenamefont {Medvedev}, \citenamefont {van~de Kruijs}, \citenamefont
  {Yakshin}, \citenamefont {Louis},\ and\ \citenamefont {Bijkerk}}]{Huang2017}%
  \BibitemOpen
  \bibfield  {author} {\bibinfo {author} {\bibfnamefont {Q.}~\bibnamefont
  {Huang}}, \bibinfo {author} {\bibfnamefont {V.}~\bibnamefont {Medvedev}},
  \bibinfo {author} {\bibfnamefont {R.}~\bibnamefont {van~de Kruijs}}, \bibinfo
  {author} {\bibfnamefont {A.}~\bibnamefont {Yakshin}}, \bibinfo {author}
  {\bibfnamefont {E.}~\bibnamefont {Louis}},\ and\ \bibinfo {author}
  {\bibfnamefont {F.}~\bibnamefont {Bijkerk}},\ }\bibfield  {title} {\bibinfo
  {title} {{Spectral tailoring of nanoscale EUV and soft x-ray multilayer
  optics}},\ }\href {https://doi.org/10.1063/1.4978290} {\bibfield  {journal}
  {\bibinfo  {journal} {Appl. Phys. Rev.}\ }\textbf {\bibinfo {volume} {4}},\
  \bibinfo {pages} {011104} (\bibinfo {year} {2017})}\BibitemShut {NoStop}%
\bibitem [{\citenamefont {Nishihara}\ \emph {et~al.}(2008)\citenamefont
  {Nishihara}, \citenamefont {Sunahara}, \citenamefont {Sasaki}, \citenamefont
  {Nunami}, \citenamefont {Tanuma}, \citenamefont {Fujioka}, \citenamefont
  {Shimada}, \citenamefont {Fujima}, \citenamefont {Furukawa}, \citenamefont
  {Kato}, \citenamefont {Koike}, \citenamefont {More}, \citenamefont
  {Murakami}, \citenamefont {Nishikawa}, \citenamefont {Zhakhovskii},
  \citenamefont {Gamata}, \citenamefont {Takata}, \citenamefont {Ueda},
  \citenamefont {Nishimura}, \citenamefont {Izawa}, \citenamefont {Miyanaga},\
  and\ \citenamefont {Mima}}]{nishihara2008plasma}%
  \BibitemOpen
  \bibfield  {author} {\bibinfo {author} {\bibfnamefont {K.}~\bibnamefont
  {Nishihara}}, \bibinfo {author} {\bibfnamefont {A.}~\bibnamefont {Sunahara}},
  \bibinfo {author} {\bibfnamefont {A.}~\bibnamefont {Sasaki}}, \bibinfo
  {author} {\bibfnamefont {M.}~\bibnamefont {Nunami}}, \bibinfo {author}
  {\bibfnamefont {H.}~\bibnamefont {Tanuma}}, \bibinfo {author} {\bibfnamefont
  {S.}~\bibnamefont {Fujioka}}, \bibinfo {author} {\bibfnamefont
  {Y.}~\bibnamefont {Shimada}}, \bibinfo {author} {\bibfnamefont
  {K.}~\bibnamefont {Fujima}}, \bibinfo {author} {\bibfnamefont
  {H.}~\bibnamefont {Furukawa}}, \bibinfo {author} {\bibfnamefont
  {T.}~\bibnamefont {Kato}}, \bibinfo {author} {\bibfnamefont {F.}~\bibnamefont
  {Koike}}, \bibinfo {author} {\bibfnamefont {R.}~\bibnamefont {More}},
  \bibinfo {author} {\bibfnamefont {M.}~\bibnamefont {Murakami}}, \bibinfo
  {author} {\bibfnamefont {T.}~\bibnamefont {Nishikawa}}, \bibinfo {author}
  {\bibfnamefont {V.}~\bibnamefont {Zhakhovskii}}, \bibinfo {author}
  {\bibfnamefont {K.}~\bibnamefont {Gamata}}, \bibinfo {author} {\bibfnamefont
  {A.}~\bibnamefont {Takata}}, \bibinfo {author} {\bibfnamefont
  {H.}~\bibnamefont {Ueda}}, \bibinfo {author} {\bibfnamefont {H.}~\bibnamefont
  {Nishimura}}, \bibinfo {author} {\bibfnamefont {Y.}~\bibnamefont {Izawa}},
  \bibinfo {author} {\bibfnamefont {N.}~\bibnamefont {Miyanaga}},\ and\
  \bibinfo {author} {\bibfnamefont {K.}~\bibnamefont {Mima}},\ }\bibfield
  {title} {\bibinfo {title} {{Plasma physics and radiation hydrodynamics in
  developing an extreme ultraviolet light source for lithography}},\ }\href
  {https://doi.org/10.1063/1.2907154} {\bibfield  {journal} {\bibinfo
  {journal} {Phys. Plasmas}\ }\textbf {\bibinfo {volume} {15}},\ \bibinfo
  {pages} {056708} (\bibinfo {year} {2008})}\BibitemShut {NoStop}%
\bibitem [{\citenamefont {Basko}(2016)}]{Basko2016}%
  \BibitemOpen
  \bibfield  {author} {\bibinfo {author} {\bibfnamefont {M.~M.}\ \bibnamefont
  {Basko}},\ }\bibfield  {title} {\bibinfo {title} {On the maximum conversion
  efficiency into the 13.5-nm extreme ultraviolet emission under a steady-state
  laser ablation of tin microspheres},\ }\href
  {https://doi.org/10.1063/1.4960684} {\bibfield  {journal} {\bibinfo
  {journal} {Phys. Plasmas}\ }\textbf {\bibinfo {volume} {23}},\ \bibinfo
  {pages} {083114} (\bibinfo {year} {2016})}\BibitemShut {NoStop}%
\bibitem [{\citenamefont {Fomenkov}\ \emph {et~al.}(2017)\citenamefont
  {Fomenkov}, \citenamefont {Brandt}, \citenamefont {Ershov}, \citenamefont
  {Schafgans}, \citenamefont {Tao}, \citenamefont {Vaschenko}, \citenamefont
  {Rokitski}, \citenamefont {Kats}, \citenamefont {Vargas}, \citenamefont
  {Purvis}, \citenamefont {Rafac}, \citenamefont {La~Fontaine}, \citenamefont
  {De~Dea}, \citenamefont {LaForge}, \citenamefont {Stewart}, \citenamefont
  {Chang}, \citenamefont {Graham}, \citenamefont {Riggs}, \citenamefont
  {Taylor}, \citenamefont {Abraham},\ and\ \citenamefont
  {Brown}}]{Fomenkov2017}%
  \BibitemOpen
  \bibfield  {author} {\bibinfo {author} {\bibfnamefont {I.}~\bibnamefont
  {Fomenkov}}, \bibinfo {author} {\bibfnamefont {D.}~\bibnamefont {Brandt}},
  \bibinfo {author} {\bibfnamefont {A.}~\bibnamefont {Ershov}}, \bibinfo
  {author} {\bibfnamefont {A.}~\bibnamefont {Schafgans}}, \bibinfo {author}
  {\bibfnamefont {Y.}~\bibnamefont {Tao}}, \bibinfo {author} {\bibfnamefont
  {G.}~\bibnamefont {Vaschenko}}, \bibinfo {author} {\bibfnamefont
  {S.}~\bibnamefont {Rokitski}}, \bibinfo {author} {\bibfnamefont
  {M.}~\bibnamefont {Kats}}, \bibinfo {author} {\bibfnamefont {M.}~\bibnamefont
  {Vargas}}, \bibinfo {author} {\bibfnamefont {M.}~\bibnamefont {Purvis}},
  \bibinfo {author} {\bibfnamefont {R.}~\bibnamefont {Rafac}}, \bibinfo
  {author} {\bibfnamefont {B.}~\bibnamefont {La~Fontaine}}, \bibinfo {author}
  {\bibfnamefont {S.}~\bibnamefont {De~Dea}}, \bibinfo {author} {\bibfnamefont
  {A.}~\bibnamefont {LaForge}}, \bibinfo {author} {\bibfnamefont
  {J.}~\bibnamefont {Stewart}}, \bibinfo {author} {\bibfnamefont
  {S.}~\bibnamefont {Chang}}, \bibinfo {author} {\bibfnamefont
  {M.}~\bibnamefont {Graham}}, \bibinfo {author} {\bibfnamefont
  {D.}~\bibnamefont {Riggs}}, \bibinfo {author} {\bibfnamefont
  {T.}~\bibnamefont {Taylor}}, \bibinfo {author} {\bibfnamefont
  {M.}~\bibnamefont {Abraham}},\ and\ \bibinfo {author} {\bibfnamefont
  {D.}~\bibnamefont {Brown}},\ }\bibfield  {title} {\bibinfo {title} {{Light
  sources for high-volume manufacturing EUV lithography: technology,
  performance, and power scaling}},\ }\href
  {https://doi.org/10.1515/aot-2017-0029} {\bibfield  {journal} {\bibinfo
  {journal} {Adv. Opt. Techn.}\ }\textbf {\bibinfo {volume} {6}},\ \bibinfo
  {pages} {173} (\bibinfo {year} {2017})}\BibitemShut {NoStop}%
\bibitem [{\citenamefont {Fujioka}\ \emph {et~al.}(2008)\citenamefont
  {Fujioka}, \citenamefont {Shimomura}, \citenamefont {Shimada}, \citenamefont
  {Maeda}, \citenamefont {Sakaguchi}, \citenamefont {Nakai}, \citenamefont
  {Aota}, \citenamefont {Nishimura}, \citenamefont {Ozaki}, \citenamefont
  {Sunahara}, \citenamefont {Nishihara}, \citenamefont {Miyanaga},
  \citenamefont {Izawa},\ and\ \citenamefont {Mima}}]{Fujioka2008}%
  \BibitemOpen
  \bibfield  {author} {\bibinfo {author} {\bibfnamefont {S.}~\bibnamefont
  {Fujioka}}, \bibinfo {author} {\bibfnamefont {M.}~\bibnamefont {Shimomura}},
  \bibinfo {author} {\bibfnamefont {Y.}~\bibnamefont {Shimada}}, \bibinfo
  {author} {\bibfnamefont {S.}~\bibnamefont {Maeda}}, \bibinfo {author}
  {\bibfnamefont {H.}~\bibnamefont {Sakaguchi}}, \bibinfo {author}
  {\bibfnamefont {Y.}~\bibnamefont {Nakai}}, \bibinfo {author} {\bibfnamefont
  {T.}~\bibnamefont {Aota}}, \bibinfo {author} {\bibfnamefont {H.}~\bibnamefont
  {Nishimura}}, \bibinfo {author} {\bibfnamefont {N.}~\bibnamefont {Ozaki}},
  \bibinfo {author} {\bibfnamefont {A.}~\bibnamefont {Sunahara}}, \bibinfo
  {author} {\bibfnamefont {K.}~\bibnamefont {Nishihara}}, \bibinfo {author}
  {\bibfnamefont {N.}~\bibnamefont {Miyanaga}}, \bibinfo {author}
  {\bibfnamefont {Y.}~\bibnamefont {Izawa}},\ and\ \bibinfo {author}
  {\bibfnamefont {K.}~\bibnamefont {Mima}},\ }\bibfield  {title} {\bibinfo
  {title} {{Pure-tin microdroplets irradiated with double laser pulses for
  efficient and minimum-mass extreme-ultraviolet light source production}},\
  }\href {https://doi.org/10.1063/1.2948874} {\bibfield  {journal} {\bibinfo
  {journal} {Appl. Phys. Lett.}\ }\textbf {\bibinfo {volume} {92}},\ \bibinfo
  {pages} {241502} (\bibinfo {year} {2008})}\BibitemShut {NoStop}%
\bibitem [{\citenamefont {Langer}\ \emph {et~al.}()\citenamefont {Langer},
  \citenamefont {Scott}, \citenamefont {Galvin}, \citenamefont {Link},
  \citenamefont {Regan},\ and\ \citenamefont {Siders}}]{Langer2020litho}%
  \BibitemOpen
  \bibfield  {author} {\bibinfo {author} {\bibfnamefont {S.}~\bibnamefont
  {Langer}}, \bibinfo {author} {\bibfnamefont {H.}~\bibnamefont {Scott}},
  \bibinfo {author} {\bibfnamefont {T.}~\bibnamefont {Galvin}}, \bibinfo
  {author} {\bibfnamefont {E.}~\bibnamefont {Link}}, \bibinfo {author}
  {\bibfnamefont {B.}~\bibnamefont {Regan}},\ and\ \bibinfo {author}
  {\bibfnamefont {C.}~\bibnamefont {Siders}},\ }\href@noop {} {\bibinfo {title}
  {{Simulations of Laser Driven EUV Sources -- the Impact of Laser
  Wavelength}}},\ \bibinfo {note}
  {\href{https://www.osti.gov/servlets/purl/1633515}{EUVL Workshop 2020,
  presented: June 11, 2020}}\BibitemShut {NoStop}%
\bibitem [{\citenamefont {Schupp}\ \emph {et~al.}(2019)\citenamefont {Schupp},
  \citenamefont {Torretti}, \citenamefont {Meijer}, \citenamefont {Bayraktar},
  \citenamefont {Sheil}, \citenamefont {Scheers}, \citenamefont {Kurilo\-vich},
  \citenamefont {Bayerle}, \citenamefont {Schafgans}, \citenamefont {Purvis},
  \citenamefont {Eikema}, \citenamefont {Witte}, \citenamefont {Ubachs},
  \citenamefont {Hoekstra},\ and\ \citenamefont {Versolato}}]{Schupp2019b}%
  \BibitemOpen
  \bibfield  {author} {\bibinfo {author} {\bibfnamefont {R.}~\bibnamefont
  {Schupp}}, \bibinfo {author} {\bibfnamefont {F.}~\bibnamefont {Torretti}},
  \bibinfo {author} {\bibfnamefont {R.~A.}\ \bibnamefont {Meijer}}, \bibinfo
  {author} {\bibfnamefont {M.}~\bibnamefont {Bayraktar}}, \bibinfo {author}
  {\bibfnamefont {J.}~\bibnamefont {Sheil}}, \bibinfo {author} {\bibfnamefont
  {J.}~\bibnamefont {Scheers}}, \bibinfo {author} {\bibfnamefont
  {D.}~\bibnamefont {Kurilo\-vich}}, \bibinfo {author} {\bibfnamefont
  {A.}~\bibnamefont {Bayerle}}, \bibinfo {author} {\bibfnamefont {A.~A.}\
  \bibnamefont {Schafgans}}, \bibinfo {author} {\bibfnamefont {M.}~\bibnamefont
  {Purvis}}, \bibinfo {author} {\bibfnamefont {K.~S.~E.}\ \bibnamefont
  {Eikema}}, \bibinfo {author} {\bibfnamefont {S.}~\bibnamefont {Witte}},
  \bibinfo {author} {\bibfnamefont {W.}~\bibnamefont {Ubachs}}, \bibinfo
  {author} {\bibfnamefont {R.}~\bibnamefont {Hoekstra}},\ and\ \bibinfo
  {author} {\bibfnamefont {O.~O.}\ \bibnamefont {Versolato}},\ }\bibfield
  {title} {\bibinfo {title} {{Radiation transport and scaling of optical depth
  in Nd:YAG laser-produced microdroplet-tin plasma}},\ }\href
  {https://doi.org/10.1063/1.5117504} {\bibfield  {journal} {\bibinfo
  {journal} {Appl. Phys. Lett.}\ }\textbf {\bibinfo {volume} {115}},\ \bibinfo
  {pages} {124101} (\bibinfo {year} {2019})}\BibitemShut {NoStop}%
\bibitem [{\citenamefont {Schupp}\ \emph {et~al.}()\citenamefont {Schupp},
  \citenamefont {Behnke}, \citenamefont {Sheil}, \citenamefont {Bouza},
  \citenamefont {Bayraktar}, \citenamefont {Ubachs}, \citenamefont {Hoekstra},\
  and\ \citenamefont {Versolato}}]{Schupp2020}%
  \BibitemOpen
  \bibfield  {author} {\bibinfo {author} {\bibfnamefont {R.}~\bibnamefont
  {Schupp}}, \bibinfo {author} {\bibfnamefont {L.}~\bibnamefont {Behnke}},
  \bibinfo {author} {\bibfnamefont {J.}~\bibnamefont {Sheil}}, \bibinfo
  {author} {\bibfnamefont {Z.}~\bibnamefont {Bouza}}, \bibinfo {author}
  {\bibfnamefont {M.}~\bibnamefont {Bayraktar}}, \bibinfo {author}
  {\bibfnamefont {W.}~\bibnamefont {Ubachs}}, \bibinfo {author} {\bibfnamefont
  {R.}~\bibnamefont {Hoekstra}},\ and\ \bibinfo {author} {\bibfnamefont
  {O.~O.}\ \bibnamefont {Versolato}},\ }\bibfield  {title} {\bibinfo {title}
  {{Characterization of 1- and 2-\textmu m-wavelength laser-produced
  microdroplet-tin plasma for generating extreme ultraviolet light}},\
  }\href@noop {} {\bibinfo  {journal} {submitted}\ }\BibitemShut {NoStop}%
\bibitem [{\citenamefont {{R. Schupp, F. Torretti, R. A. Meijer, M. Bayraktar,
  J. Scheers, D. Kurilo\-vich, A. Bayerle, K. S. E. Eikema, S. Witte, W.
  Ubachs, R. Hoekstra and O. O. Versolato}}(2019)}]{Schupp2019}%
  \BibitemOpen
\bibfield  {journal} {  }\bibfield  {author} {\bibinfo {author} {\bibnamefont
  {{R. Schupp, F. Torretti, R. A. Meijer, M. Bayraktar, J. Scheers, D.
  Kurilo\-vich, A. Bayerle, K. S. E. Eikema, S. Witte, W. Ubachs, R. Hoekstra
  and O. O. Versolato}}},\ }\bibfield  {title} {\bibinfo {title} {{Efficient
  generation of extreme ultraviolet light from Nd: YAG-driven microdroplet-tin
  plasma}},\ }\href
  {https://doi.org/https://doi.org/10.1103/PhysRevApplied.12.014010} {\bibfield
   {journal} {\bibinfo  {journal} {Phys. Rev. Appl.}\ }\textbf {\bibinfo
  {volume} {12}},\ \bibinfo {pages} {014010} (\bibinfo {year}
  {2019})}\BibitemShut {NoStop}%
\bibitem [{\citenamefont {Fujioka}\ \emph {et~al.}(2005)\citenamefont
  {Fujioka}, \citenamefont {Nishimura}, \citenamefont {Nishihara},
  \citenamefont {Sasaki}, \citenamefont {Sunahara}, \citenamefont {Okuno},
  \citenamefont {Ueda}, \citenamefont {Ando}, \citenamefont {Tao},
  \citenamefont {Shimada}, \citenamefont {Hashimoto}, \citenamefont {Yamaura},
  \citenamefont {Shigemori}, \citenamefont {Nakai}, \citenamefont {Nagai},
  \citenamefont {Norimatsu}, \citenamefont {Nishikawa}, \citenamefont
  {Miyanaga}, \citenamefont {Izawa},\ and\ \citenamefont
  {Mima}}]{Fujioka2005opacity}%
  \BibitemOpen
  \bibfield  {author} {\bibinfo {author} {\bibfnamefont {S.}~\bibnamefont
  {Fujioka}}, \bibinfo {author} {\bibfnamefont {H.}~\bibnamefont {Nishimura}},
  \bibinfo {author} {\bibfnamefont {K.}~\bibnamefont {Nishihara}}, \bibinfo
  {author} {\bibfnamefont {A.}~\bibnamefont {Sasaki}}, \bibinfo {author}
  {\bibfnamefont {A.}~\bibnamefont {Sunahara}}, \bibinfo {author}
  {\bibfnamefont {T.}~\bibnamefont {Okuno}}, \bibinfo {author} {\bibfnamefont
  {N.}~\bibnamefont {Ueda}}, \bibinfo {author} {\bibfnamefont {T.}~\bibnamefont
  {Ando}}, \bibinfo {author} {\bibfnamefont {Y.}~\bibnamefont {Tao}}, \bibinfo
  {author} {\bibfnamefont {Y.}~\bibnamefont {Shimada}}, \bibinfo {author}
  {\bibfnamefont {K.}~\bibnamefont {Hashimoto}}, \bibinfo {author}
  {\bibfnamefont {M.}~\bibnamefont {Yamaura}}, \bibinfo {author} {\bibfnamefont
  {K.}~\bibnamefont {Shigemori}}, \bibinfo {author} {\bibfnamefont
  {M.}~\bibnamefont {Nakai}}, \bibinfo {author} {\bibfnamefont
  {K.}~\bibnamefont {Nagai}}, \bibinfo {author} {\bibfnamefont
  {T.}~\bibnamefont {Norimatsu}}, \bibinfo {author} {\bibfnamefont
  {T.}~\bibnamefont {Nishikawa}}, \bibinfo {author} {\bibfnamefont
  {N.}~\bibnamefont {Miyanaga}}, \bibinfo {author} {\bibfnamefont
  {Y.}~\bibnamefont {Izawa}},\ and\ \bibinfo {author} {\bibfnamefont
  {K.}~\bibnamefont {Mima}},\ }\bibfield  {title} {\bibinfo {title} {{Opacity
  Effect on Extreme Ultraviolet Radiation from Laser-Produced Tin Plasmas}},\
  }\href {https://doi.org/10.1103/PhysRevLett.95.235004} {\bibfield  {journal}
  {\bibinfo  {journal} {Phys. Rev. Lett.}\ }\textbf {\bibinfo {volume} {95}},\
  \bibinfo {pages} {235004} (\bibinfo {year} {2005})}\BibitemShut {NoStop}%
\bibitem [{\citenamefont {Arisholm}\ \emph {et~al.}(2004)\citenamefont
  {Arisholm}, \citenamefont {Nordseth},\ and\ \citenamefont
  {Rustad}}]{Arisholm2004}%
  \BibitemOpen
  \bibfield  {author} {\bibinfo {author} {\bibfnamefont {G.}~\bibnamefont
  {Arisholm}}, \bibinfo {author} {\bibfnamefont {{\O}.}~\bibnamefont
  {Nordseth}},\ and\ \bibinfo {author} {\bibfnamefont {G.}~\bibnamefont
  {Rustad}},\ }\bibfield  {title} {\bibinfo {title} {{Optical parametric master
  oscillator and power amplifier for efficient conversion of high-energy pulses
  with high beam quality}},\ }\href {https://doi.org/10.1364/OPEX.12.004189}
  {\bibfield  {journal} {\bibinfo  {journal} {Opt. Express}\ }\textbf {\bibinfo
  {volume} {12}},\ \bibinfo {pages} {4189} (\bibinfo {year}
  {2004})}\BibitemShut {NoStop}%
\bibitem [{\citenamefont {Behnke}\ \emph {et~al.}(2021)\citenamefont {Behnke},
  \citenamefont {Schupp}, \citenamefont {Bouza}, \citenamefont {Bayraktar},
  \citenamefont {Mazzotta}, \citenamefont {Meijer}, \citenamefont {Sheil},
  \citenamefont {Witte}, \citenamefont {Ubachs}, \citenamefont {Hoekstra},\
  and\ \citenamefont {Versolato}}]{Behnke2021}%
  \BibitemOpen
  \bibfield  {author} {\bibinfo {author} {\bibfnamefont {L.}~\bibnamefont
  {Behnke}}, \bibinfo {author} {\bibfnamefont {R.}~\bibnamefont {Schupp}},
  \bibinfo {author} {\bibfnamefont {Z.}~\bibnamefont {Bouza}}, \bibinfo
  {author} {\bibfnamefont {M.}~\bibnamefont {Bayraktar}}, \bibinfo {author}
  {\bibfnamefont {Z.}~\bibnamefont {Mazzotta}}, \bibinfo {author}
  {\bibfnamefont {R.}~\bibnamefont {Meijer}}, \bibinfo {author} {\bibfnamefont
  {J.}~\bibnamefont {Sheil}}, \bibinfo {author} {\bibfnamefont
  {S.}~\bibnamefont {Witte}}, \bibinfo {author} {\bibfnamefont
  {W.}~\bibnamefont {Ubachs}}, \bibinfo {author} {\bibfnamefont
  {R.}~\bibnamefont {Hoekstra}},\ and\ \bibinfo {author} {\bibfnamefont
  {O.~O.}\ \bibnamefont {Versolato}},\ }\bibfield  {title} {\bibinfo {title}
  {{Extreme ultraviolet light from a tin plasma driven by a 2-\textmu
  m-wavelength laser}},\ }\href {https://doi.org/10.1364/OE.411539} {\bibfield
  {journal} {\bibinfo  {journal} {Opt. Express}\ }\textbf {\bibinfo {volume}
  {29}},\ \bibinfo {pages} {4475} (\bibinfo {year} {2021})}\BibitemShut
  {NoStop}%
\bibitem [{\citenamefont {Danson}\ \emph {et~al.}(2019)\citenamefont {Danson},
  \citenamefont {Haefner}, \citenamefont {Bromage}, \citenamefont {Butcher},
  \citenamefont {Chanteloup}, \citenamefont {Chowdhury}, \citenamefont
  {Galvanauskas}, \citenamefont {Gizzi}, \citenamefont {Hein}, \citenamefont
  {Hillier}, \citenamefont {Hopps}, \citenamefont {Kato}, \citenamefont
  {Khazanov}, \citenamefont {Kodama}, \citenamefont {Korn}, \citenamefont {Li},
  \citenamefont {Li}, \citenamefont {Limpert}, \citenamefont {Ma},
  \citenamefont {Chang-hee}, \citenamefont {Neely}, \citenamefont
  {Papadopoulos}, \citenamefont {Penman}, \citenamefont {Qian}, \citenamefont
  {Rocca}, \citenamefont {Shaykin}, \citenamefont {Siders}, \citenamefont
  {Spindloe}, \citenamefont {Szatmári}, \citenamefont {Trines}, \citenamefont
  {Zhu}, \citenamefont {Zhu},\ and\ \citenamefont
  {Zuegel}}]{Danson2019petawatt}%
  \BibitemOpen
  \bibfield  {author} {\bibinfo {author} {\bibfnamefont {C.~N.}\ \bibnamefont
  {Danson}}, \bibinfo {author} {\bibfnamefont {C.}~\bibnamefont {Haefner}},
  \bibinfo {author} {\bibfnamefont {J.}~\bibnamefont {Bromage}}, \bibinfo
  {author} {\bibfnamefont {T.}~\bibnamefont {Butcher}}, \bibinfo {author}
  {\bibfnamefont {J.-C.~F.}\ \bibnamefont {Chanteloup}}, \bibinfo {author}
  {\bibfnamefont {E.~A.}\ \bibnamefont {Chowdhury}}, \bibinfo {author}
  {\bibfnamefont {A.}~\bibnamefont {Galvanauskas}}, \bibinfo {author}
  {\bibfnamefont {L.~A.}\ \bibnamefont {Gizzi}}, \bibinfo {author}
  {\bibfnamefont {J.}~\bibnamefont {Hein}}, \bibinfo {author} {\bibfnamefont
  {D.~I.}\ \bibnamefont {Hillier}}, \bibinfo {author} {\bibfnamefont {N.~W.}\
  \bibnamefont {Hopps}}, \bibinfo {author} {\bibfnamefont {Y.}~\bibnamefont
  {Kato}}, \bibinfo {author} {\bibfnamefont {E.~A.}\ \bibnamefont {Khazanov}},
  \bibinfo {author} {\bibfnamefont {R.}~\bibnamefont {Kodama}}, \bibinfo
  {author} {\bibfnamefont {G.}~\bibnamefont {Korn}}, \bibinfo {author}
  {\bibfnamefont {R.}~\bibnamefont {Li}}, \bibinfo {author} {\bibfnamefont
  {Y.}~\bibnamefont {Li}}, \bibinfo {author} {\bibfnamefont {J.}~\bibnamefont
  {Limpert}}, \bibinfo {author} {\bibfnamefont {J.}~\bibnamefont {Ma}},
  \bibinfo {author} {\bibfnamefont {N.}~\bibnamefont {Chang-hee}}, \bibinfo
  {author} {\bibfnamefont {D.}~\bibnamefont {Neely}}, \bibinfo {author}
  {\bibfnamefont {D.}~\bibnamefont {Papadopoulos}}, \bibinfo {author}
  {\bibfnamefont {R.~R.}\ \bibnamefont {Penman}}, \bibinfo {author}
  {\bibfnamefont {L.}~\bibnamefont {Qian}}, \bibinfo {author} {\bibfnamefont
  {J.~J.}\ \bibnamefont {Rocca}}, \bibinfo {author} {\bibfnamefont {A.~A.}\
  \bibnamefont {Shaykin}}, \bibinfo {author} {\bibfnamefont {C.~W.}\
  \bibnamefont {Siders}}, \bibinfo {author} {\bibfnamefont {C.}~\bibnamefont
  {Spindloe}}, \bibinfo {author} {\bibfnamefont {S.}~\bibnamefont {Szatmári}},
  \bibinfo {author} {\bibfnamefont {R.~M. G.~M.}\ \bibnamefont {Trines}},
  \bibinfo {author} {\bibfnamefont {J.}~\bibnamefont {Zhu}}, \bibinfo {author}
  {\bibfnamefont {P.}~\bibnamefont {Zhu}},\ and\ \bibinfo {author}
  {\bibfnamefont {J.~D.}\ \bibnamefont {Zuegel}},\ }\bibfield  {title}
  {\bibinfo {title} {{Petawatt and exawatt class lasers worldwide}},\ }\href
  {https://doi.org/10.1017/hpl.2019.36} {\bibfield  {journal} {\bibinfo
  {journal} {High Power Laser Sci. Eng.}\ }\textbf {\bibinfo {volume} {7}}
  (\bibinfo {year} {2019})}\BibitemShut {NoStop}%
\bibitem [{\citenamefont {Sistrunk}\ \emph {et~al.}(2019)\citenamefont
  {Sistrunk}, \citenamefont {Alessi}, \citenamefont {Bayramian}, \citenamefont
  {Chesnut}, \citenamefont {Erlandson}, \citenamefont {Galvin}, \citenamefont
  {Gibson}, \citenamefont {Nguyen}, \citenamefont {Reagan}, \citenamefont
  {Schaffers}, \citenamefont {Siders}, \citenamefont {Spinka},\ and\
  \citenamefont {Haefner}}]{Sistrunk2019}%
  \BibitemOpen
  \bibfield  {author} {\bibinfo {author} {\bibfnamefont {E.}~\bibnamefont
  {Sistrunk}}, \bibinfo {author} {\bibfnamefont {D.~A.}\ \bibnamefont
  {Alessi}}, \bibinfo {author} {\bibfnamefont {A.}~\bibnamefont {Bayramian}},
  \bibinfo {author} {\bibfnamefont {K.}~\bibnamefont {Chesnut}}, \bibinfo
  {author} {\bibfnamefont {A.}~\bibnamefont {Erlandson}}, \bibinfo {author}
  {\bibfnamefont {T.~C.}\ \bibnamefont {Galvin}}, \bibinfo {author}
  {\bibfnamefont {D.}~\bibnamefont {Gibson}}, \bibinfo {author} {\bibfnamefont
  {H.}~\bibnamefont {Nguyen}}, \bibinfo {author} {\bibfnamefont
  {B.}~\bibnamefont {Reagan}}, \bibinfo {author} {\bibfnamefont
  {K.}~\bibnamefont {Schaffers}}, \bibinfo {author} {\bibfnamefont {C.~W.}\
  \bibnamefont {Siders}}, \bibinfo {author} {\bibfnamefont {T.}~\bibnamefont
  {Spinka}},\ and\ \bibinfo {author} {\bibfnamefont {C.}~\bibnamefont
  {Haefner}},\ }\bibfield  {title} {\bibinfo {title} {{Laser Technology
  Development for High Peak Power Lasers Achieving Kilowatt Average Power and
  Beyond}},\ }in\ \href {https://doi.org/10.1117/12.2525380} {\emph {\bibinfo
  {booktitle} {{Short-pulse High-energy Lasers and Ultrafast Optical
  Technologies}}}},\ Vol.\ \bibinfo {volume} {11034},\ \bibinfo {editor}
  {edited by\ \bibinfo {editor} {\bibfnamefont {P.}~\bibnamefont {Bakule}}\
  and\ \bibinfo {editor} {\bibfnamefont {C.~L.}\ \bibnamefont {Haefner}}}\
  (\bibinfo  {publisher} {SPIE},\ \bibinfo {year} {2019})\ pp.\ \bibinfo
  {pages} {1--8}\BibitemShut {NoStop}%
\bibitem [{\citenamefont {Matsukuma}\ \emph {et~al.}(2015)\citenamefont
  {Matsukuma}, \citenamefont {Sunahara}, \citenamefont {Yanagida},
  \citenamefont {Tomuro}, \citenamefont {Kouge}, \citenamefont {Kodama},
  \citenamefont {Hosoda}, \citenamefont {Fujioka},\ and\ \citenamefont
  {Nishimura}}]{Matsukuma2015}%
  \BibitemOpen
  \bibfield  {author} {\bibinfo {author} {\bibfnamefont {H.}~\bibnamefont
  {Matsukuma}}, \bibinfo {author} {\bibfnamefont {A.}~\bibnamefont {Sunahara}},
  \bibinfo {author} {\bibfnamefont {T.}~\bibnamefont {Yanagida}}, \bibinfo
  {author} {\bibfnamefont {H.}~\bibnamefont {Tomuro}}, \bibinfo {author}
  {\bibfnamefont {K.}~\bibnamefont {Kouge}}, \bibinfo {author} {\bibfnamefont
  {T.}~\bibnamefont {Kodama}}, \bibinfo {author} {\bibfnamefont
  {T.}~\bibnamefont {Hosoda}}, \bibinfo {author} {\bibfnamefont
  {S.}~\bibnamefont {Fujioka}},\ and\ \bibinfo {author} {\bibfnamefont
  {H.}~\bibnamefont {Nishimura}},\ }\bibfield  {title} {\bibinfo {title}
  {{Correlation between laser absorption and radiation conversion efficiency in
  laser produced tin plasma}},\ }\href {https://doi.org/10.1063/1.4931698}
  {\bibfield  {journal} {\bibinfo  {journal} {Appl. Phys. Lett.}\ }\textbf
  {\bibinfo {volume} {107}},\ \bibinfo {pages} {121103} (\bibinfo {year}
  {2015})}\BibitemShut {NoStop}%
\bibitem [{\citenamefont {Higashiguchi}\ \emph {et~al.}(2006)\citenamefont
  {Higashiguchi}, \citenamefont {Dojyo}, \citenamefont {Hamada}, \citenamefont
  {Sasaki},\ and\ \citenamefont {Kubodera}}]{Higashiguchi2006}%
  \BibitemOpen
  \bibfield  {author} {\bibinfo {author} {\bibfnamefont {T.}~\bibnamefont
  {Higashiguchi}}, \bibinfo {author} {\bibfnamefont {N.}~\bibnamefont {Dojyo}},
  \bibinfo {author} {\bibfnamefont {M.}~\bibnamefont {Hamada}}, \bibinfo
  {author} {\bibfnamefont {W.}~\bibnamefont {Sasaki}},\ and\ \bibinfo {author}
  {\bibfnamefont {S.}~\bibnamefont {Kubodera}},\ }\bibfield  {title} {\bibinfo
  {title} {{Low-debris, efficient laser-produced plasma extreme ultraviolet
  source by use of a regenerative liquid microjet target containing tin dioxide
  (SnO$_2$) nanoparticles}},\ }\href {https://doi.org/10.1063/1.2206131}
  {\bibfield  {journal} {\bibinfo  {journal} {Appl. Phys. Lett.}\ }\textbf
  {\bibinfo {volume} {88}},\ \bibinfo {pages} {201503} (\bibinfo {year}
  {2006})}\BibitemShut {NoStop}%
\bibitem [{\citenamefont {Nakamura}\ \emph {et~al.}(2008)\citenamefont
  {Nakamura}, \citenamefont {Akiyama}, \citenamefont {Okazaki}, \citenamefont
  {Tamaru}, \citenamefont {Takahashi},\ and\ \citenamefont
  {Okada}}]{Nakamura2008}%
  \BibitemOpen
  \bibfield  {author} {\bibinfo {author} {\bibfnamefont {D.}~\bibnamefont
  {Nakamura}}, \bibinfo {author} {\bibfnamefont {T.}~\bibnamefont {Akiyama}},
  \bibinfo {author} {\bibfnamefont {K.}~\bibnamefont {Okazaki}}, \bibinfo
  {author} {\bibfnamefont {K.}~\bibnamefont {Tamaru}}, \bibinfo {author}
  {\bibfnamefont {A.}~\bibnamefont {Takahashi}},\ and\ \bibinfo {author}
  {\bibfnamefont {T.}~\bibnamefont {Okada}},\ }\bibfield  {title} {\bibinfo
  {title} {{Ablation dynamics of tin micro-droplet irradiated by double pulse
  laser used for extreme ultraviolet lithography source}},\ }\href
  {https://doi.org/10.1088/0022-3727/41/24/245210} {\bibfield  {journal}
  {\bibinfo  {journal} {J. Phys. D: Appl. Phys.}\ }\textbf {\bibinfo {volume}
  {41}},\ \bibinfo {pages} {245210} (\bibinfo {year} {2008})}\BibitemShut
  {NoStop}%
\bibitem [{\citenamefont {Giovannini}\ and\ \citenamefont
  {Abhari}(2013)}]{Giovannini2013}%
  \BibitemOpen
  \bibfield  {author} {\bibinfo {author} {\bibfnamefont {A.~Z.}\ \bibnamefont
  {Giovannini}}\ and\ \bibinfo {author} {\bibfnamefont {R.~S.}\ \bibnamefont
  {Abhari}},\ }\bibfield  {title} {\bibinfo {title} {{Three-dimensional extreme
  ultraviolet emission from a droplet-based laser-produced plasma}},\ }\href
  {https://doi.org/10.1063/1.4815955} {\bibfield  {journal} {\bibinfo
  {journal} {J. Appl. Phys.}\ }\textbf {\bibinfo {volume} {114}},\ \bibinfo
  {pages} {033303} (\bibinfo {year} {2013})}\BibitemShut {NoStop}%
\bibitem [{\citenamefont {Sequoia}\ \emph {et~al.}(2008)\citenamefont
  {Sequoia}, \citenamefont {Tao}, \citenamefont {Yuspeh}, \citenamefont
  {Burdt},\ and\ \citenamefont {Tillack}}]{Sequoia2008}%
  \BibitemOpen
  \bibfield  {author} {\bibinfo {author} {\bibfnamefont {K.~L.}\ \bibnamefont
  {Sequoia}}, \bibinfo {author} {\bibfnamefont {Y.}~\bibnamefont {Tao}},
  \bibinfo {author} {\bibfnamefont {S.}~\bibnamefont {Yuspeh}}, \bibinfo
  {author} {\bibfnamefont {R.}~\bibnamefont {Burdt}},\ and\ \bibinfo {author}
  {\bibfnamefont {M.~S.}\ \bibnamefont {Tillack}},\ }\bibfield  {title}
  {\bibinfo {title} {{Two dimensional expansion effects on angular distribution
  of 13.5 nm in-band extreme ultraviolet emission from laser-produced Sn
  plasma}},\ }\href {https://doi.org/10.1063/1.2938717} {\bibfield  {journal}
  {\bibinfo  {journal} {Appl. Phys. Lett.}\ }\textbf {\bibinfo {volume} {92}},\
  \bibinfo {pages} {221505} (\bibinfo {year} {2008})}\BibitemShut {NoStop}%
\bibitem [{\citenamefont {Shimada}\ \emph {et~al.}(2005)\citenamefont
  {Shimada}, \citenamefont {Nishimura}, \citenamefont {Nakai}, \citenamefont
  {Hashimoto}, \citenamefont {Yamaura}, \citenamefont {Tao}, \citenamefont
  {Shigemori}, \citenamefont {Okuno}, \citenamefont {Nishihara}, \citenamefont
  {Kawamura}, \citenamefont {Sunahara}, \citenamefont {Nishikawa},
  \citenamefont {Sasaki}, \citenamefont {Nagai}, \citenamefont {Norimatsu},
  \citenamefont {Fujioka}, \citenamefont {Uchida}, \citenamefont {Miyanaga},
  \citenamefont {Izawa},\ and\ \citenamefont {Yamanaka}}]{Shimada2005}%
  \BibitemOpen
  \bibfield  {author} {\bibinfo {author} {\bibfnamefont {Y.}~\bibnamefont
  {Shimada}}, \bibinfo {author} {\bibfnamefont {H.}~\bibnamefont {Nishimura}},
  \bibinfo {author} {\bibfnamefont {M.}~\bibnamefont {Nakai}}, \bibinfo
  {author} {\bibfnamefont {K.}~\bibnamefont {Hashimoto}}, \bibinfo {author}
  {\bibfnamefont {M.}~\bibnamefont {Yamaura}}, \bibinfo {author} {\bibfnamefont
  {Y.}~\bibnamefont {Tao}}, \bibinfo {author} {\bibfnamefont {K.}~\bibnamefont
  {Shigemori}}, \bibinfo {author} {\bibfnamefont {T.}~\bibnamefont {Okuno}},
  \bibinfo {author} {\bibfnamefont {K.}~\bibnamefont {Nishihara}}, \bibinfo
  {author} {\bibfnamefont {T.}~\bibnamefont {Kawamura}}, \bibinfo {author}
  {\bibfnamefont {A.}~\bibnamefont {Sunahara}}, \bibinfo {author}
  {\bibfnamefont {T.}~\bibnamefont {Nishikawa}}, \bibinfo {author}
  {\bibfnamefont {A.}~\bibnamefont {Sasaki}}, \bibinfo {author} {\bibfnamefont
  {K.}~\bibnamefont {Nagai}}, \bibinfo {author} {\bibfnamefont
  {T.}~\bibnamefont {Norimatsu}}, \bibinfo {author} {\bibfnamefont
  {S.}~\bibnamefont {Fujioka}}, \bibinfo {author} {\bibfnamefont
  {S.}~\bibnamefont {Uchida}}, \bibinfo {author} {\bibfnamefont
  {N.}~\bibnamefont {Miyanaga}}, \bibinfo {author} {\bibfnamefont
  {Y.}~\bibnamefont {Izawa}},\ and\ \bibinfo {author} {\bibfnamefont
  {C.}~\bibnamefont {Yamanaka}},\ }\bibfield  {title} {\bibinfo {title}
  {{Characterization of extreme ultraviolet emission from laser-produced
  spherical tin plasma generated with multiple laser beams}},\ }\href
  {https://doi.org/10.1063/1.1856697} {\bibfield  {journal} {\bibinfo
  {journal} {Appl. Phys. Lett.}\ }\textbf {\bibinfo {volume} {86}},\ \bibinfo
  {pages} {051501} (\bibinfo {year} {2005})}\BibitemShut {NoStop}%
\bibitem [{\citenamefont {Ando}\ \emph {et~al.}(2006)\citenamefont {Ando},
  \citenamefont {Fujioka}, \citenamefont {Nishimura}, \citenamefont {Ueda},
  \citenamefont {Yasuda}, \citenamefont {Nagai}, \citenamefont {Norimatsu},
  \citenamefont {Murakami}, \citenamefont {Nishihara}, \citenamefont
  {Miyanaga}, \citenamefont {Izawa}, \citenamefont {Mima},\ and\ \citenamefont
  {Sunahara}}]{Ando2006optimum}%
  \BibitemOpen
  \bibfield  {author} {\bibinfo {author} {\bibfnamefont {T.}~\bibnamefont
  {Ando}}, \bibinfo {author} {\bibfnamefont {S.}~\bibnamefont {Fujioka}},
  \bibinfo {author} {\bibfnamefont {H.}~\bibnamefont {Nishimura}}, \bibinfo
  {author} {\bibfnamefont {N.}~\bibnamefont {Ueda}}, \bibinfo {author}
  {\bibfnamefont {Y.}~\bibnamefont {Yasuda}}, \bibinfo {author} {\bibfnamefont
  {K.}~\bibnamefont {Nagai}}, \bibinfo {author} {\bibfnamefont
  {T.}~\bibnamefont {Norimatsu}}, \bibinfo {author} {\bibfnamefont
  {M.}~\bibnamefont {Murakami}}, \bibinfo {author} {\bibfnamefont
  {K.}~\bibnamefont {Nishihara}}, \bibinfo {author} {\bibfnamefont
  {N.}~\bibnamefont {Miyanaga}}, \bibinfo {author} {\bibfnamefont
  {Y.}~\bibnamefont {Izawa}}, \bibinfo {author} {\bibfnamefont
  {K.}~\bibnamefont {Mima}},\ and\ \bibinfo {author} {\bibfnamefont
  {A.}~\bibnamefont {Sunahara}},\ }\bibfield  {title} {\bibinfo {title}
  {{Optimum laser pulse duration for efficient extreme ultraviolet light
  generation from laser-produced tin plasmas}},\ }\href
  {https://doi.org/10.1063/1.2361260} {\bibfield  {journal} {\bibinfo
  {journal} {Appl. Phys. Lett.}\ }\textbf {\bibinfo {volume} {89}},\ \bibinfo
  {pages} {151501} (\bibinfo {year} {2006})}\BibitemShut {NoStop}%
\bibitem [{\citenamefont {Choi}\ \emph {et~al.}(2000)\citenamefont {Choi},
  \citenamefont {Daido}, \citenamefont {Yamagami}, \citenamefont {Nagai},
  \citenamefont {Norimatsu}, \citenamefont {Takabe}, \citenamefont {Suzuki},
  \citenamefont {Nakayama},\ and\ \citenamefont {Matsui}}]{Choi2000}%
  \BibitemOpen
  \bibfield  {author} {\bibinfo {author} {\bibfnamefont {I.~W.}\ \bibnamefont
  {Choi}}, \bibinfo {author} {\bibfnamefont {H.}~\bibnamefont {Daido}},
  \bibinfo {author} {\bibfnamefont {S.}~\bibnamefont {Yamagami}}, \bibinfo
  {author} {\bibfnamefont {K.}~\bibnamefont {Nagai}}, \bibinfo {author}
  {\bibfnamefont {T.}~\bibnamefont {Norimatsu}}, \bibinfo {author}
  {\bibfnamefont {H.}~\bibnamefont {Takabe}}, \bibinfo {author} {\bibfnamefont
  {M.}~\bibnamefont {Suzuki}}, \bibinfo {author} {\bibfnamefont
  {T.}~\bibnamefont {Nakayama}},\ and\ \bibinfo {author} {\bibfnamefont
  {T.}~\bibnamefont {Matsui}},\ }\bibfield  {title} {\bibinfo {title}
  {{Detailed space-resolved characterization of a laser-plasma soft-x-ray
  source at 13.5-nm wavelength with tin and its oxides}},\ }\href
  {https://doi.org/10.1364/JOSAB.17.001616} {\bibfield  {journal} {\bibinfo
  {journal} {J. Opt. Soc. Am. B}\ }\textbf {\bibinfo {volume} {17}},\ \bibinfo
  {pages} {1616} (\bibinfo {year} {2000})}\BibitemShut {NoStop}%
\bibitem [{\citenamefont {Freeman}\ \emph {et~al.}(2012)\citenamefont
  {Freeman}, \citenamefont {Harilal}, \citenamefont {Verhoff}, \citenamefont
  {Hassanein},\ and\ \citenamefont {Rice}}]{Freeman2012laser}%
  \BibitemOpen
  \bibfield  {author} {\bibinfo {author} {\bibfnamefont {J.~R.}\ \bibnamefont
  {Freeman}}, \bibinfo {author} {\bibfnamefont {S.~S.}\ \bibnamefont
  {Harilal}}, \bibinfo {author} {\bibfnamefont {B.}~\bibnamefont {Verhoff}},
  \bibinfo {author} {\bibfnamefont {A.}~\bibnamefont {Hassanein}},\ and\
  \bibinfo {author} {\bibfnamefont {B.}~\bibnamefont {Rice}},\ }\bibfield
  {title} {\bibinfo {title} {{Laser wavelength dependence on angular emission
  dynamics of Nd:YAG laser-produced Sn plasmas}},\ }\href
  {https://doi.org/10.1088/0963-0252/21/5/055003} {\bibfield  {journal}
  {\bibinfo  {journal} {Plasma Sources Sci. Technol.}\ }\textbf {\bibinfo
  {volume} {21}},\ \bibinfo {pages} {055003} (\bibinfo {year}
  {2012})}\BibitemShut {NoStop}%
\bibitem [{\citenamefont {Morris}\ \emph {et~al.}(2008)\citenamefont {Morris},
  \citenamefont {O'Reilly}, \citenamefont {Dunne},\ and\ \citenamefont
  {Hayden}}]{Morris2008Angular}%
  \BibitemOpen
  \bibfield  {author} {\bibinfo {author} {\bibfnamefont {O.}~\bibnamefont
  {Morris}}, \bibinfo {author} {\bibfnamefont {F.}~\bibnamefont {O'Reilly}},
  \bibinfo {author} {\bibfnamefont {P.}~\bibnamefont {Dunne}},\ and\ \bibinfo
  {author} {\bibfnamefont {P.}~\bibnamefont {Hayden}},\ }\bibfield  {title}
  {\bibinfo {title} {{Angular emission and self-absorption studies of a tin
  laser produced plasma extreme ultraviolet source between 10 and 18\,nm}},\
  }\href {https://doi.org/10.1063/1.2945645} {\bibfield  {journal} {\bibinfo
  {journal} {Appl. Phys. Lett.}\ }\textbf {\bibinfo {volume} {92}},\ \bibinfo
  {pages} {231503} (\bibinfo {year} {2008})}\BibitemShut {NoStop}%
\bibitem [{\citenamefont {Kurilovich}\ \emph {et~al.}(2016)\citenamefont
  {Kurilovich}, \citenamefont {Klein}, \citenamefont {Torretti}, \citenamefont
  {Lassise}, \citenamefont {Hoekstra}, \citenamefont {Ubachs}, \citenamefont
  {Gelderblom},\ and\ \citenamefont {Versolato}}]{Kurilovich2016}%
  \BibitemOpen
  \bibfield  {author} {\bibinfo {author} {\bibfnamefont {D.}~\bibnamefont
  {Kurilovich}}, \bibinfo {author} {\bibfnamefont {A.~L.}\ \bibnamefont
  {Klein}}, \bibinfo {author} {\bibfnamefont {F.}~\bibnamefont {Torretti}},
  \bibinfo {author} {\bibfnamefont {A.}~\bibnamefont {Lassise}}, \bibinfo
  {author} {\bibfnamefont {R.}~\bibnamefont {Hoekstra}}, \bibinfo {author}
  {\bibfnamefont {W.}~\bibnamefont {Ubachs}}, \bibinfo {author} {\bibfnamefont
  {H.}~\bibnamefont {Gelderblom}},\ and\ \bibinfo {author} {\bibfnamefont
  {O.~O.}\ \bibnamefont {Versolato}},\ }\bibfield  {title} {\bibinfo {title}
  {Plasma propulsion of a metallic microdroplet and its deformation upon laser
  impact},\ }\href {https://doi.org/10.1103/PhysRevApplied.6.014018} {\bibfield
   {journal} {\bibinfo  {journal} {Phys. Rev. Appl.}\ }\textbf {\bibinfo
  {volume} {6}},\ \bibinfo {pages} {014018} (\bibinfo {year}
  {2016})}\BibitemShut {NoStop}%
\bibitem [{\citenamefont {Kurilovich}\ \emph {et~al.}(2018)\citenamefont
  {Kurilovich}, \citenamefont {Basko}, \citenamefont {Kim}, \citenamefont
  {Torretti}, \citenamefont {Schupp}, \citenamefont {Visschers}, \citenamefont
  {Scheers}, \citenamefont {Hoekstra}, \citenamefont {Ubachs},\ and\
  \citenamefont {Versolato}}]{Kurilovich2018}%
  \BibitemOpen
  \bibfield  {author} {\bibinfo {author} {\bibfnamefont {D.}~\bibnamefont
  {Kurilovich}}, \bibinfo {author} {\bibfnamefont {M.~M.}\ \bibnamefont
  {Basko}}, \bibinfo {author} {\bibfnamefont {D.~A.}\ \bibnamefont {Kim}},
  \bibinfo {author} {\bibfnamefont {F.}~\bibnamefont {Torretti}}, \bibinfo
  {author} {\bibfnamefont {R.}~\bibnamefont {Schupp}}, \bibinfo {author}
  {\bibfnamefont {J.~C.}\ \bibnamefont {Visschers}}, \bibinfo {author}
  {\bibfnamefont {J.}~\bibnamefont {Scheers}}, \bibinfo {author} {\bibfnamefont
  {R.}~\bibnamefont {Hoekstra}}, \bibinfo {author} {\bibfnamefont
  {W.}~\bibnamefont {Ubachs}},\ and\ \bibinfo {author} {\bibfnamefont {O.~O.}\
  \bibnamefont {Versolato}},\ }\bibfield  {title} {\bibinfo {title} {{Power-law
  scaling of plasma pressure on laser-ablated tin microdroplets}},\ }\href
  {https://doi.org/10.1063/1.5010899} {\bibfield  {journal} {\bibinfo
  {journal} {Phys. Plasmas}\ }\textbf {\bibinfo {volume} {25}},\ \bibinfo
  {pages} {012709} (\bibinfo {year} {2018})}\BibitemShut {NoStop}%
\bibitem [{\citenamefont {Liu}\ \emph {et~al.}(2020)\citenamefont {Liu},
  \citenamefont {Kurilovich}, \citenamefont {Gelderblom},\ and\ \citenamefont
  {Versolato}}]{Liu2020a}%
  \BibitemOpen
  \bibfield  {author} {\bibinfo {author} {\bibfnamefont {B.}~\bibnamefont
  {Liu}}, \bibinfo {author} {\bibfnamefont {D.}~\bibnamefont {Kurilovich}},
  \bibinfo {author} {\bibfnamefont {H.}~\bibnamefont {Gelderblom}},\ and\
  \bibinfo {author} {\bibfnamefont {O.~O.}\ \bibnamefont {Versolato}},\
  }\bibfield  {title} {\bibinfo {title} {{Mass Loss from a Stretching
  Semitransparent Sheet of Liquid Tin}},\ }\href
  {https://doi.org/10.1103/PhysRevApplied.13.024035} {\bibfield  {journal}
  {\bibinfo  {journal} {Phys. Rev. Appl.}\ }\textbf {\bibinfo {volume} {13}},\
  \bibinfo {pages} {024035} (\bibinfo {year} {2020})}\BibitemShut {NoStop}%
\bibitem [{\citenamefont {Liu}\ \emph {et~al.}(2021)\citenamefont {Liu},
  \citenamefont {Meijer}, \citenamefont {Hernandez-Rueda}, \citenamefont
  {Kurilovich}, \citenamefont {Mazzotta}, \citenamefont {Witte},\ and\
  \citenamefont {Versolato}}]{Liu2021a}%
  \BibitemOpen
  \bibfield  {author} {\bibinfo {author} {\bibfnamefont {B.}~\bibnamefont
  {Liu}}, \bibinfo {author} {\bibfnamefont {R.~A.}\ \bibnamefont {Meijer}},
  \bibinfo {author} {\bibfnamefont {J.}~\bibnamefont {Hernandez-Rueda}},
  \bibinfo {author} {\bibfnamefont {D.}~\bibnamefont {Kurilovich}}, \bibinfo
  {author} {\bibfnamefont {Z.}~\bibnamefont {Mazzotta}}, \bibinfo {author}
  {\bibfnamefont {S.}~\bibnamefont {Witte}},\ and\ \bibinfo {author}
  {\bibfnamefont {O.~O.}\ \bibnamefont {Versolato}},\ }\bibfield  {title}
  {\bibinfo {title} {{Laser-induced vaporization of a stretching sheet of
  liquid tin}},\ }\href@noop {} {\bibfield  {journal} {\bibinfo  {journal}
  {Journal of Applied Physics}\ }\textbf {\bibinfo {volume} {129}},\ \bibinfo
  {pages} {053302} (\bibinfo {year} {2021})}\BibitemShut {NoStop}%
\bibitem [{\citenamefont {de~Faria~Pinto}\ \emph {et~al.}(2021)\citenamefont
  {de~Faria~Pinto}, \citenamefont {Mathijssen}, \citenamefont {Meijer},
  \citenamefont {Zhang}, \citenamefont {Bayerle}, \citenamefont {Kurilovich},
  \citenamefont {Versolato}, \citenamefont {Eikema},\ and\ \citenamefont
  {Witte}}]{Pinto2021}%
  \BibitemOpen
  \bibfield  {author} {\bibinfo {author} {\bibfnamefont {T.}~\bibnamefont
  {de~Faria~Pinto}}, \bibinfo {author} {\bibfnamefont {J.}~\bibnamefont
  {Mathijssen}}, \bibinfo {author} {\bibfnamefont {R.}~\bibnamefont {Meijer}},
  \bibinfo {author} {\bibfnamefont {H.}~\bibnamefont {Zhang}}, \bibinfo
  {author} {\bibfnamefont {A.}~\bibnamefont {Bayerle}}, \bibinfo {author}
  {\bibfnamefont {D.}~\bibnamefont {Kurilovich}}, \bibinfo {author}
  {\bibfnamefont {O.~O.}\ \bibnamefont {Versolato}}, \bibinfo {author}
  {\bibfnamefont {K.~S.~E.}\ \bibnamefont {Eikema}},\ and\ \bibinfo {author}
  {\bibfnamefont {S.}~\bibnamefont {Witte}},\ }\bibfield  {title} {\bibinfo
  {title} {{Cylindrically and non-cylindrically symmetric expansion dynamics of
  tin microdroplets after ultrashort laser pulse impact}},\ }\href@noop {}
  {\bibfield  {journal} {\bibinfo  {journal} {Appl. Phys. A}\ }\textbf
  {\bibinfo {volume} {127}},\ \bibinfo {pages} {1} (\bibinfo {year}
  {2021})}\BibitemShut {NoStop}%
\bibitem [{\citenamefont {Meijer}\ \emph {et~al.}(2017)\citenamefont {Meijer},
  \citenamefont {Stodolna}, \citenamefont {Eikema},\ and\ \citenamefont
  {Witte}}]{Meijer2017}%
  \BibitemOpen
  \bibfield  {author} {\bibinfo {author} {\bibfnamefont {R.~A.}\ \bibnamefont
  {Meijer}}, \bibinfo {author} {\bibfnamefont {A.~S.}\ \bibnamefont
  {Stodolna}}, \bibinfo {author} {\bibfnamefont {K.~S.~E.}\ \bibnamefont
  {Eikema}},\ and\ \bibinfo {author} {\bibfnamefont {S.}~\bibnamefont
  {Witte}},\ }\bibfield  {title} {\bibinfo {title} {{High-energy Nd:YAG laser
  system with arbitrary sub-nanosecond pulse shaping capability}},\ }\href
  {https://doi.org/10.1364/OL.42.002758} {\bibfield  {journal} {\bibinfo
  {journal} {Opt. Lett.}\ }\textbf {\bibinfo {volume} {42}},\ \bibinfo {pages}
  {2758} (\bibinfo {year} {2017})}\BibitemShut {NoStop}%
\bibitem [{\citenamefont {Bayraktar}\ \emph {et~al.}(2016)\citenamefont
  {Bayraktar}, \citenamefont {Bastiaens}, \citenamefont {Bruineman},
  \citenamefont {Vratzov},\ and\ \citenamefont
  {Bijkerk}}]{Bayraktar2016broadband}%
  \BibitemOpen
  \bibfield  {author} {\bibinfo {author} {\bibfnamefont {M.}~\bibnamefont
  {Bayraktar}}, \bibinfo {author} {\bibfnamefont {H.~M.}\ \bibnamefont
  {Bastiaens}}, \bibinfo {author} {\bibfnamefont {C.}~\bibnamefont
  {Bruineman}}, \bibinfo {author} {\bibfnamefont {B.}~\bibnamefont {Vratzov}},\
  and\ \bibinfo {author} {\bibfnamefont {F.}~\bibnamefont {Bijkerk}},\
  }\bibfield  {title} {\bibinfo {title} {{Broadband transmission grating
  spectrometer for measuring the emission spectrum of EUV sources}},\ }\href
  {https://ris.utwente.nl/ws/files/6488314/nevac2016-bayraktar.pdf} {\bibfield
  {journal} {\bibinfo  {journal} {NEVAC blad}\ }\textbf {\bibinfo {volume}
  {54}},\ \bibinfo {pages} {14} (\bibinfo {year} {2016})}\BibitemShut {NoStop}%
\bibitem [{\citenamefont {Yamaura}\ \emph {et~al.}(2005)\citenamefont
  {Yamaura}, \citenamefont {Uchida}, \citenamefont {Sunahara}, \citenamefont
  {Shimada}, \citenamefont {Nishimura}, \citenamefont {Fujioka}, \citenamefont
  {Okuno}, \citenamefont {Hashimoto}, \citenamefont {Nagai}, \citenamefont
  {Norimatsu}, \citenamefont {Nishihara}, \citenamefont {Miyanga},
  \citenamefont {Izawa},\ and\ \citenamefont {Yamanaka}}]{Yamaura2005}%
  \BibitemOpen
  \bibfield  {author} {\bibinfo {author} {\bibfnamefont {M.}~\bibnamefont
  {Yamaura}}, \bibinfo {author} {\bibfnamefont {S.}~\bibnamefont {Uchida}},
  \bibinfo {author} {\bibfnamefont {A.}~\bibnamefont {Sunahara}}, \bibinfo
  {author} {\bibfnamefont {Y.}~\bibnamefont {Shimada}}, \bibinfo {author}
  {\bibfnamefont {H.}~\bibnamefont {Nishimura}}, \bibinfo {author}
  {\bibfnamefont {S.}~\bibnamefont {Fujioka}}, \bibinfo {author} {\bibfnamefont
  {T.}~\bibnamefont {Okuno}}, \bibinfo {author} {\bibfnamefont
  {K.}~\bibnamefont {Hashimoto}}, \bibinfo {author} {\bibfnamefont
  {K.}~\bibnamefont {Nagai}}, \bibinfo {author} {\bibfnamefont
  {T.}~\bibnamefont {Norimatsu}}, \bibinfo {author} {\bibfnamefont
  {K.}~\bibnamefont {Nishihara}}, \bibinfo {author} {\bibfnamefont
  {N.}~\bibnamefont {Miyanga}}, \bibinfo {author} {\bibfnamefont
  {Y.}~\bibnamefont {Izawa}},\ and\ \bibinfo {author} {\bibfnamefont
  {C.}~\bibnamefont {Yamanaka}},\ }\bibfield  {title} {\bibinfo {title}
  {{Characterization of extreme ultraviolet emission using the fourth harmonic
  of a Nd:YAG laser}},\ }\href {https://doi.org/10.1063/1.1915507} {\bibfield
  {journal} {\bibinfo  {journal} {Appl. Phys. Lett.}\ }\textbf {\bibinfo
  {volume} {86}},\ \bibinfo {pages} {181107} (\bibinfo {year}
  {2005})}\BibitemShut {NoStop}%
\bibitem [{\citenamefont {Hayden}\ \emph {et~al.}(2006)\citenamefont {Hayden},
  \citenamefont {Cummings}, \citenamefont {Murphy}, \citenamefont {O'Sullivan},
  \citenamefont {Sheridan}, \citenamefont {White},\ and\ \citenamefont
  {Dunne}}]{Hayden2006}%
  \BibitemOpen
  \bibfield  {author} {\bibinfo {author} {\bibfnamefont {P.}~\bibnamefont
  {Hayden}}, \bibinfo {author} {\bibfnamefont {A.}~\bibnamefont {Cummings}},
  \bibinfo {author} {\bibfnamefont {N.}~\bibnamefont {Murphy}}, \bibinfo
  {author} {\bibfnamefont {G.}~\bibnamefont {O'Sullivan}}, \bibinfo {author}
  {\bibfnamefont {P.}~\bibnamefont {Sheridan}}, \bibinfo {author}
  {\bibfnamefont {J.}~\bibnamefont {White}},\ and\ \bibinfo {author}
  {\bibfnamefont {P.}~\bibnamefont {Dunne}},\ }\bibfield  {title} {\bibinfo
  {title} {{13.5\,nm extreme ultraviolet emission from tin based laser produced
  plasma sources}},\ }\href {https://doi.org/10.1063/1.2191477} {\bibfield
  {journal} {\bibinfo  {journal} {J. Appl. Phys.}\ }\textbf {\bibinfo {volume}
  {99}},\ \bibinfo {pages} {093302} (\bibinfo {year} {2006})}\BibitemShut
  {NoStop}%
\bibitem [{\citenamefont {Yuspeh}\ \emph {et~al.}(2008)\citenamefont {Yuspeh},
  \citenamefont {Sequoia}, \citenamefont {Tao}, \citenamefont {Tillack},
  \citenamefont {Burdt},\ and\ \citenamefont
  {Najmabadi}}]{Yuspeh2008optimization}%
  \BibitemOpen
  \bibfield  {author} {\bibinfo {author} {\bibfnamefont {S.}~\bibnamefont
  {Yuspeh}}, \bibinfo {author} {\bibfnamefont {K.~L.}\ \bibnamefont {Sequoia}},
  \bibinfo {author} {\bibfnamefont {Y.}~\bibnamefont {Tao}}, \bibinfo {author}
  {\bibfnamefont {M.~S.}\ \bibnamefont {Tillack}}, \bibinfo {author}
  {\bibfnamefont {R.}~\bibnamefont {Burdt}},\ and\ \bibinfo {author}
  {\bibfnamefont {F.}~\bibnamefont {Najmabadi}},\ }\bibfield  {title} {\bibinfo
  {title} {{Optimization of the size ratio of Sn sphere and laser focal spot
  for an extreme ultraviolet light source}},\ }\href
  {https://doi.org/10.1063/1.3036956} {\bibfield  {journal} {\bibinfo
  {journal} {Appl. Phys. Lett.}\ }\textbf {\bibinfo {volume} {93}},\ \bibinfo
  {pages} {221503} (\bibinfo {year} {2008})}\BibitemShut {NoStop}%
\bibitem [{\citenamefont {Torretti}\ \emph {et~al.}(2018)\citenamefont
  {Torretti}, \citenamefont {Schupp}, \citenamefont {Kurilovich}, \citenamefont
  {Bayerle}, \citenamefont {Scheers}, \citenamefont {Ubachs}, \citenamefont
  {Hoekstra},\ and\ \citenamefont {Versolato}}]{Torretti2018}%
  \BibitemOpen
  \bibfield  {author} {\bibinfo {author} {\bibfnamefont {F.}~\bibnamefont
  {Torretti}}, \bibinfo {author} {\bibfnamefont {R.}~\bibnamefont {Schupp}},
  \bibinfo {author} {\bibfnamefont {D.}~\bibnamefont {Kurilovich}}, \bibinfo
  {author} {\bibfnamefont {A.}~\bibnamefont {Bayerle}}, \bibinfo {author}
  {\bibfnamefont {J.}~\bibnamefont {Scheers}}, \bibinfo {author} {\bibfnamefont
  {W.}~\bibnamefont {Ubachs}}, \bibinfo {author} {\bibfnamefont
  {R.}~\bibnamefont {Hoekstra}},\ and\ \bibinfo {author} {\bibfnamefont
  {O.~O.}\ \bibnamefont {Versolato}},\ }\bibfield  {title} {\bibinfo {title}
  {{Short-wavelength out-of-band EUV emission from Sn laser-produced plasma}},\
  }\href {https://doi.org/10.1088/1361-6455/aaa593} {\bibfield  {journal}
  {\bibinfo  {journal} {J. Phys. B: At. Mol. Opt. Phys.}\ }\textbf {\bibinfo
  {volume} {51}},\ \bibinfo {pages} {045005} (\bibinfo {year}
  {2018})}\BibitemShut {NoStop}%
\bibitem [{\citenamefont {O'Sullivan}\ and\ \citenamefont
  {Faulkner}(1994)}]{OSullivan1994}%
  \BibitemOpen
  \bibfield  {author} {\bibinfo {author} {\bibfnamefont {G.~D.}\ \bibnamefont
  {O'Sullivan}}\ and\ \bibinfo {author} {\bibfnamefont {R.}~\bibnamefont
  {Faulkner}},\ }\bibfield  {title} {\bibinfo {title} {{Tunable narrowband soft
  x-ray source for projection lithography}},\ }\href
  {https://doi.org/10.1117/12.186840} {\bibfield  {journal} {\bibinfo
  {journal} {Opt. Eng.}\ }\textbf {\bibinfo {volume} {33}},\ \bibinfo {pages}
  {3978} (\bibinfo {year} {1994})}\BibitemShut {NoStop}%
\bibitem [{\citenamefont {van~de Kerkhof}\ \emph {et~al.}(2020)\citenamefont
  {van~de Kerkhof}, \citenamefont {Liu}, \citenamefont {Meeuwissen},
  \citenamefont {Zhang}, \citenamefont {Bayraktar}, \citenamefont {de~Kruif},\
  and\ \citenamefont {Davydova}}]{Kerkhof2020}%
  \BibitemOpen
  \bibfield  {author} {\bibinfo {author} {\bibfnamefont {M.}~\bibnamefont
  {van~de Kerkhof}}, \bibinfo {author} {\bibfnamefont {F.}~\bibnamefont {Liu}},
  \bibinfo {author} {\bibfnamefont {M.}~\bibnamefont {Meeuwissen}}, \bibinfo
  {author} {\bibfnamefont {X.}~\bibnamefont {Zhang}}, \bibinfo {author}
  {\bibfnamefont {M.}~\bibnamefont {Bayraktar}}, \bibinfo {author}
  {\bibfnamefont {R.}~\bibnamefont {de~Kruif}},\ and\ \bibinfo {author}
  {\bibfnamefont {N.}~\bibnamefont {Davydova}},\ }\bibfield  {title} {\bibinfo
  {title} {{High-power EUV lithography: spectral purity and imaging
  performance}},\ }\href {https://doi.org/10.1117/1.JMM.19.3.033801} {\bibfield
   {journal} {\bibinfo  {journal} {J. Micro Nanolithogr. MEMS MOEMS}\ }\textbf
  {\bibinfo {volume} {19}},\ \bibinfo {pages} {033801} (\bibinfo {year}
  {2020})}\BibitemShut {NoStop}%
\bibitem [{\citenamefont {Reijers}\ \emph {et~al.}(2018)\citenamefont
  {Reijers}, \citenamefont {Kurilovich}, \citenamefont {Torretti},
  \citenamefont {Gelderblom},\ and\ \citenamefont
  {Versolato}}]{Reijers2018laser}%
  \BibitemOpen
  \bibfield  {author} {\bibinfo {author} {\bibfnamefont {S.~A.}\ \bibnamefont
  {Reijers}}, \bibinfo {author} {\bibfnamefont {D.}~\bibnamefont {Kurilovich}},
  \bibinfo {author} {\bibfnamefont {F.}~\bibnamefont {Torretti}}, \bibinfo
  {author} {\bibfnamefont {H.}~\bibnamefont {Gelderblom}},\ and\ \bibinfo
  {author} {\bibfnamefont {O.~O.}\ \bibnamefont {Versolato}},\ }\bibfield
  {title} {\bibinfo {title} {{Laser-to-droplet alignment sensitivity relevant
  for laser-produced plasma sources of extreme ultraviolet light}},\
  }\href@noop {} {\bibfield  {journal} {\bibinfo  {journal} {J. Appl. Phys.}\
  }\textbf {\bibinfo {volume} {124}},\ \bibinfo {pages} {013102} (\bibinfo
  {year} {2018})}\BibitemShut {NoStop}%
\bibitem [{\citenamefont {Sizyuk}\ and\ \citenamefont
  {Hassanein}(2020)}]{sizyuk2020tuning}%
  \BibitemOpen
  \bibfield  {author} {\bibinfo {author} {\bibfnamefont {T.}~\bibnamefont
  {Sizyuk}}\ and\ \bibinfo {author} {\bibfnamefont {A.}~\bibnamefont
  {Hassanein}},\ }\bibfield  {title} {\bibinfo {title} {{Tuning laser
  wavelength and pulse duration to improve the conversion efficiency and
  performance of EUV sources for nanolithography}},\ }\href
  {https://doi.org/10.1063/5.0018576} {\bibfield  {journal} {\bibinfo
  {journal} {Phys. Plasmas}\ }\textbf {\bibinfo {volume} {27}},\ \bibinfo
  {pages} {103507} (\bibinfo {year} {2020})}\BibitemShut {NoStop}%
\bibitem [{\citenamefont {Chen}\ \emph {et~al.}(2015)\citenamefont {Chen},
  \citenamefont {Wang}, \citenamefont {Duan}, \citenamefont {Lan},
  \citenamefont {Chen}, \citenamefont {Zuo},\ and\ \citenamefont
  {Lu}}]{chen2015angular}%
  \BibitemOpen
  \bibfield  {author} {\bibinfo {author} {\bibfnamefont {H.}~\bibnamefont
  {Chen}}, \bibinfo {author} {\bibfnamefont {X.}~\bibnamefont {Wang}}, \bibinfo
  {author} {\bibfnamefont {L.}~\bibnamefont {Duan}}, \bibinfo {author}
  {\bibfnamefont {H.}~\bibnamefont {Lan}}, \bibinfo {author} {\bibfnamefont
  {Z.}~\bibnamefont {Chen}}, \bibinfo {author} {\bibfnamefont {D.}~\bibnamefont
  {Zuo}},\ and\ \bibinfo {author} {\bibfnamefont {P.}~\bibnamefont {Lu}},\
  }\bibfield  {title} {\bibinfo {title} {{Angular distribution of ions and
  extreme ultraviolet emission in laser-produced tin droplet plasma}},\ }\href
  {https://doi.org/10.1063/1.4921532} {\bibfield  {journal} {\bibinfo
  {journal} {J. Appl. Phys.}\ }\textbf {\bibinfo {volume} {117}},\ \bibinfo
  {pages} {193302} (\bibinfo {year} {2015})}\BibitemShut {NoStop}%
\bibitem [{\citenamefont {Apruzese}\ \emph {et~al.}(2002)\citenamefont
  {Apruzese}, \citenamefont {Davis}, \citenamefont {Whitney}, \citenamefont
  {Thornhill}, \citenamefont {Kepple}, \citenamefont {Clark}, \citenamefont
  {Deeney}, \citenamefont {Coverdale},\ and\ \citenamefont
  {Sanford}}]{Apruzese2002}%
  \BibitemOpen
  \bibfield  {author} {\bibinfo {author} {\bibfnamefont {J.~P.}\ \bibnamefont
  {Apruzese}}, \bibinfo {author} {\bibfnamefont {J.}~\bibnamefont {Davis}},
  \bibinfo {author} {\bibfnamefont {K.~G.}\ \bibnamefont {Whitney}}, \bibinfo
  {author} {\bibfnamefont {J.~W.}\ \bibnamefont {Thornhill}}, \bibinfo {author}
  {\bibfnamefont {P.~C.}\ \bibnamefont {Kepple}}, \bibinfo {author}
  {\bibfnamefont {R.~W.}\ \bibnamefont {Clark}}, \bibinfo {author}
  {\bibfnamefont {C.}~\bibnamefont {Deeney}}, \bibinfo {author} {\bibfnamefont
  {C.~A.}\ \bibnamefont {Coverdale}},\ and\ \bibinfo {author} {\bibfnamefont
  {T.~W.~L.}\ \bibnamefont {Sanford}},\ }\bibfield  {title} {\bibinfo {title}
  {{The physics of radiation transport in dense plasmas}},\ }\href
  {http://dx.doi.org/10.1063/1.1446038} {\bibfield  {journal} {\bibinfo
  {journal} {Phys. Plasmas}\ }\textbf {\bibinfo {volume} {9}},\ \bibinfo
  {pages} {2411} (\bibinfo {year} {2002})}\BibitemShut {NoStop}%
\end{thebibliography}
%

\end{document}